\documentclass{article}
\usepackage{graphicx}
\usepackage{amsfonts}
\usepackage{amsmath}
\usepackage{amssymb}
\usepackage{url}

\usepackage{dsfont}
\usepackage{fancyhdr}
\usepackage{indentfirst}
\usepackage{enumerate}
\usepackage[normalem]{ulem}
\usepackage{mathrsfs}
\usepackage{multirow}
\usepackage{diagbox}

\usepackage{todonotes,soul}

 \usepackage[colorlinks=true,citecolor=blue]{hyperref}
\usepackage{amsthm}
\usepackage{color}

\definecolor{DarkGreen}{rgb}{0.2,0.6,0.2}

\definecolor{purple}{rgb}{0.6,0.3,0.8}

\usepackage{natbib}
\usepackage{comment}

\def\d{\mathrm{d}}

\newcommand{\R}{\mathbb{R}}

\newcommand{\p}{\mathbb{P}}
\newcommand{\X}{\mathcal{X}}

\newcommand{\id}{\mathds{1}}

\renewcommand{\ge}{\geqslant}

\renewcommand{\geq}{\geqslant}
\renewcommand{\leq}{\leqslant}
\renewcommand{\epsilon}{\varepsilon}
\newcommand{\esssup}{\mathrm{ess\mbox{-}sup}}
\newcommand{\essinf}{\mathrm{ess\mbox{-}inf}}
\renewcommand{\cdots}{\dots}
\theoremstyle{plain}
\newtheorem{theorem}{Theorem}
\newtheorem{corollary}{Corollary}

\newtheorem{proposition}{Proposition}
\theoremstyle{definition}

\newtheorem{example}{Example}

\theoremstyle{remark}

\newtheorem*{remark*}{Remark}

\newcommand{\dsquare}{\mathop{  \square} \displaylimits}
\newcommand{\VaR}{\mathrm{VaR}}

\newcommand{\ES}{\mathrm{ES}}
\newcommand{\tbl}{\textcolor{black}}

\newcommand{\thedate}{\today}

\usepackage{setspace}

\usepackage{footmisc}
\begin{document}
\title{Risk sharing with Lambda Value-at-Risk under heterogeneous beliefs}
\author{
Peng Liu\thanks{School of Mathematics, Statistics and Actuarial Science, University of Essex, UK. Email: peng.liu@essex.ac.uk} \and Andreas Tsanakas\thanks{Bayes Business School, City St George's, University of London, London, UK. Email: a.tsanakas.1@citystgeorges.ac.uk}\and Yunran Wei\thanks{School of Mathematics and Statistics, Carleton University, Canada. Email: Yunran.Wei@carleton.ca}
}

  \date{\thedate}

 \maketitle
\begin{abstract}
In this paper, we study the risk sharing problem among multiple agents using Lambda Value-at-Risk  as their preference functional, under heterogeneous beliefs, where beliefs are represented by several probability measures. We obtain semi-explicit formulas for the inf-convolution of multiple Lambda Value-at-Risk measures under heterogeneous beliefs and the explicit forms of the corresponding optimal allocations. To show the impact of belief heterogeneity, we consider three cases: homogeneous beliefs, conditional beliefs, and general beliefs with two agents. For those cases, we find more explicit expressions for the inf-convolution, showing the influence of the relation of the beliefs on the inf-convolution.  Moreover, we consider, in a two-agent setting, the inf-convolution of one Lambda Value-at-Risk and a general risk measure, including expected utility, distortion risk measures and Lambda Value-at-Risk as special cases, with differing beliefs. The expression of the inf-convolution and the form of the optimal allocation are obtained. In all above cases we demonstrate that trivial outcomes arise when both belief inconsistency and risk tolerance are high. Finally, we discuss risk sharing for an alternative definition of Lambda Value-at-Risk.

\begin{bfseries}Key-words\end{bfseries}: Lambda Value-at-Risk; Value-at-Risk; Risk sharing; Inf-convolution; Distortion risk measure; Expected shortfall; $\mathrm{CoVaR}$; $\mathrm{CoES}$
\end{abstract}

 \section{Introduction}\label{sec:1}

The Pareto-optimal risk sharing problem, with the preferences of agents represented by risk measures, has been studied extensively since the introduction of convex and coherent risk measures by \cite{ADEH99}, \cite{FS02} and \cite{FR05}. For instance, convex risk measures were used in \cite{BE05}, \cite{JST08} and \cite{FS08} to investigate the risk sharing problem, showing the existence of the optimal allocation, which is \emph{comonotonic} under the assumption of law-invariance.  Recently, more focus has been placed on the risk sharing problem for non-convex risk measures, such as quantile-based risk measures including Value-at-Risk ($\VaR$) and Expected Shortfall ($\ES$) as special cases (e.g., \cite{ELW18}, \cite{LMWW22} and \cite{W18}) and distortion riskmetrics including inter-quantile-range as an example (e.g., \cite{LLW23a}). In those papers, it is shown that the optimal allocation is pairwise countermonotonic, as defined in \cite{LLW23}.

In the above papers, the agents are assumed to share the same beliefs on the distribution of the future risk. However, in the current regulatory frameworks (e.g., \cite{BS16}), internal models are extensively used, leading to heterogeneous views of different agents regarding the same future risk. Moreover, the belief heterogeneity may also stem from the asymmetric information accessed by the agents.  We refer to \cite{E17} for a discussion of application of internal models in banking and insurance and \cite{X13} for a relevant discussion in finance. The model/belief heterogeneity in the risk sharing problem has been considered by several papers in the literature.  For instance, under model heterogeneity, the risk sharing between two agents endowed with concave  monetary utilities  was studied in \cite{AS09}, where the random variables were constrained to a finite $\sigma$-algebra and the existence of Pareto optimal allocation was discussed.  Risk sharing for $\VaR$ or $\ES$  under heterogeneous beliefs was studied in \cite{ELMW19}, who showed that optimal risk allocations are pairwise countermonotonic with model heterogeneity. \cite{L20} investigated the comonotonic risk sharing with heterogeneous beliefs for distortion risk measures and \cite{L22} studied the risk sharing for consistent risk measures with model heterogeneity, obtaining a sufficient condition for the existence of the optimal allocations under some assumptions on the relation of the beliefs.  Model heterogeneity was also studied in the context of optimal insurance and reinsurance design; see \cite{AGP15}, \cite{B16}, \cite{C19}, \cite{BG20}, \cite{ABCC21} and the references therein.

In this paper, we use Lambda Value-at-Risk ($\Lambda\VaR$) to represent the agents' preferences.  As an extension of $\VaR$, $\Lambda\VaR$  was introduced by \cite{FMP14}  by changing the fixed probability level to a probability/loss function $1-\Lambda$.  The $\Lambda$ function can be chosen either increasing (relatively risk-averse) or decreasing (relatively risk-seeking), representing decision makers' individual risk appetite, as shown in \cite{FMP14}.  The choice of $\Lambda$ function based on data was studied in  \cite{HMP18}.  Compared with $\VaR$, one advantage of Lambda Value-at-Risk is its ability to  distinguish the tail risk for decreasing $\Lambda$ functions, in the same spirit as the Loss $\VaR$ proposed in \cite{BBM20}.  For increasing $\Lambda$ functions, $\Lambda\VaR$ may incorporate some additional requirement such as  risk manager's judgement in the process of risk management; see \cite{BP22}.   Moreover, the recent literature shows that $\Lambda\VaR$ satisfies  other desirable properties.  \cite{HWWX24} showed that $\Lambda\VaR$ with increasing $\Lambda$ function satisfies quasi-star-shapedness,  which is a property weaker than quasi-convexity and penalizing a kind of risk concentration. In addition, $\Lambda\VaR$ with increasing $\Lambda$ functions also satisfies cash subadditivity, which is useful to measure the future financial loss with stochastic interest rates; see \cite{KR09} and \cite{HWWX24}. We refer to \cite{BP22} for the monotonicity, locality and  other properties, and \cite{BPR17} for robustness, elicitability and consistency. The risk sharing problem for $\Lambda\VaR$ under homogeneous beliefs was studied in \cite{L24} and \cite{XH24}, where the expressions for the inf-convolution and the forms of the optimal allocation were derived. The $\Lambda\VaR$ was also applied to optimal reinsurance design with model homogeneity in \cite{BBB23} and \cite{BCHW24}. Moreover, the application of $\Lambda\VaR$ to robust portfolio selection and sensitivity analysis can be found in \cite{HL24} and \cite{IPP22}. 

In our study, we  put our focus on the risk sharing problem with $\Lambda\VaR$ and belief heterogeneity, extending many results in the literature.  In Section \ref{sec:multiple}, we study the inf-convolution of multiple $\Lambda\VaR$ with heterogeneous beliefs, where the beliefs are represented by a set of probability measures. We obtain a semi-explicit formula for the inf-convolution and explicit forms for the optimal risk allocations, which show that agents take contingent losses on disjoint sets of the sample space.

The relation between the beliefs plays a crucial role in the risk sharing problem; see e.g., \cite{C19}. To study the impact of belief heterogeneity, we consider three cases in Section \ref{sec:multiple}: homogeneous beliefs, conditional beliefs, and general beliefs with two agents.  For homogeneous beliefs, we show that the inf-convolution of $\Lambda\VaR$ is still a $\Lambda\VaR$ for general $\Lambda$ functions, extending the results in \cite{L24} and \cite{XH24}, where the monotonicity of the $\Lambda$ functions is typically required. In particular, our results include the case that some $\Lambda\VaR$ have increasing $\Lambda$ functions and others have decreasing $\Lambda$ functions, demonstrating that the agents may have relatively different risk appetites.  The conditional beliefs reflect the asymmetric information obtained by the agents such that each agent has slightly different concern.  Under this setup, we also obtain an explicit formula for the inf-convolution of $\Lambda\VaR$, showing that the inf-convolution of $\Lambda\VaR$ under conditional beliefs is also a $\Lambda\VaR$ under a new conditional belief. Conditional beliefs are closely related to the conditional risk measures proposed in the literature to quantify systemic risk, such as  $\mathrm{CoVaR}$ (\citet{AB16} and \citet{GE13}) and
 $\mathrm{CoES}$ (\citet{MS14}). The risk sharing problem in this case can be interpreted as risk sharing with some Co-risk measures ($\mathrm{Co\Lambda\VaR}$, a new conditional risk measure proposed in Section \ref{sec:multiple}). For the third case, we consider the risk sharing between two agents. With the aid of the Lebesgue decomposition theorem, we find a more explicit formula for the inf-convolution of $\Lambda\VaR$ with general heterogeneous beliefs. Note that we also give the explicit expression for the inf-convolution of $\VaR$ with heterogeneous beliefs, which are also new to the literature.

In Section \ref{sec:5}, we study the inf-convolution of one $\Lambda\VaR$ and a general monotone risk measure with belief heterogeneity. The expression of the inf-convolution and the form of the optimal allocation are derived. Then we consider some concrete examples.  For conditional beliefs, we obtain explicit expressions for the inf-convolution of one $\Lambda\VaR$ and one distortion risk measure/$\Lambda\VaR^+$. Moreover, for general heterogeneous beliefs, we find the formula for the inf-convolution of one $\Lambda\VaR$ and one $\ES$/expected utility. In Section \ref{sec:4}, we consider the inf-convolution of one $\Lambda\VaR^+$ (an alternative definition of Lambda Value-at-Risk) and a general monotone risk measure with belief heterogeneity. We obtain the expression of the inf-convolution, which is complicated, and the forms of the optimal allocation. Our results include the inf-convolution of two $\Lambda\VaR^+$ under heterogeneous beliefs as a special case. The results here are not explicit and very different from the previous sections, as they involve the shape of the $\Lambda$ function and the best case of risk aggregation under dependence uncertainty. This means that $\Lambda\VaR^+$ is very different from $\Lambda\VaR$ for its application within the risk sharing problem. Some special cases are considered to illustrate our theory in Section \ref{sec:4}.

Throughout the paper, the stated results provide conditions for trivial, i.e., infinite, inf-convolutions. We consistently show that these conditions arise when, broadly speaking, agents' beliefs are strongly inconsistent and risk tolerances are high. Hence, our contribution illuminates how the entanglement of beliefs and preferences impacts key features of the risk sharing problem with $\Lambda\VaR$.

   The notation and key definitions are given in Section \ref{sec:2} and all the proofs are delegated to Appendix \ref{Appendix}.

\section{Notation and Definitions}\label{sec:2}

For a given probability space $(\Omega,\mathcal F,\p)$, let $L^\infty$ denote the collection of all bounded random variables. 
Moreover, let $\mathcal X$ be a set of random variables containing  $L^\infty$. We say $\X$ is unbounded if $\X$ contains unbounded random variables.  We  suppose $\X$ has good enough properties to conduct our study, such as $\mathcal X=L^p$ for some $p\geq 0$, where $L^p$ represents the set of all random variables with finite $p$-th moments. Throughout this paper, we consider the random variables themselves but not the corresponding equivalent classes, because, as will be seen in the sequel,  more than one probability measures are considered simultaneously.

For any $X\in\X$, a positive value of $X$ represents a financial loss. For a risk measure, that is, a mapping $\rho:\X\to\R$,  we say that $\rho$ \emph{monotone} if $X\leq Y$ implies $\rho(X)\leq \rho(Y)$; and $\rho$ is \emph{cash-additive} if $\rho(X+c)=\rho(X)+c$ for $X\in\X$ and $c\in\R$. We say $\rho$ is a \emph{monetary} risk measure if it is monotone and cash-additive.   Furthermore, $\rho$ is \emph{law-invariant} under a probability measure $\mathbb Q$ if  for all $X,Y\in \X$, \begin{equation} \label{eq:1}  X\overset{\mathbb Q}{=} Y ~~ \Rightarrow ~~\rho(X)=\rho(Y) , \end{equation} where $\overset{\mathbb Q}{=}$ stands for equality in distribution under  $\mathbb Q$. Note that $\mathbb Q$ may differ from $\mathbb P$. Throughout this paper, all risk measures used are law-invariant. Hence, when evaluating the risk measure $\rho$ on $X$, with respect to the measure $\mathbb Q$, we will consistently use the notation $\rho^{\mathbb Q}(X)$ for the resulting value. For more details on risk measures, one can refer to Chapter 4 of \cite{FS16}. 

For a probability measure $\mathbb Q$, the distribution function of $X$ under $\mathbb Q$ is denoted as $F_X^{\mathbb Q}$. For $X\in\X$, $F_X^{\mathbb Q, -1}$ represents its left-quantile under $\mathbb Q$, which is defined by
$$F_X^{\mathbb Q, -1}(p)=\inf\{x: F_X^{\mathbb Q}(x)\geq p\},~p\in (0,1]$$
with the convention that $\inf\emptyset=\infty$.
  For any $X\in\X$, if $\mathbb Q$ is atomless, we denote by $ U_X^{\mathbb Q}$   a uniform random variable on $[0,1]$ under $\mathbb Q$ such that $X=F_X^{\mathbb Q, -1}(U_X^{\mathbb Q})$ a.s. under $\mathbb Q$. The existence of such $U_X^{\mathbb Q}$ for any random variable $X$ is guaranteed by e.g., Lemma A.32 of \cite{FS16}.
  
\tbl{We next define tail risk measures, which are important for our results later. For a random variable $X\in \mathcal X$, $p\in (0,1]$ and an atomless probability measure $\mathbb Q$, we call 
$$X_p=(F_X^{\mathbb Q})^{-1}(1-p+pU_X^{\mathbb Q})$$
the tail risk of $X$ beyond its $(1-p)$-quantile under $\mathbb Q$.
The distribution of $X_p$ under $\mathbb Q$ is given by
$$\mathbb{Q}(X_p\leq x)=\frac{(F_X^{\mathbb Q}(x)-(1-p))_+}{p},~x\in\R,$$
where $x_+=\max(x,0)$.
We say $\rho$ is a $p$-tail risk measure under $\mathbb Q$ for some  $p\in (0,1)$ if $X_p \overset{\mathbb Q}{=}Y_p$ implies $\rho^{\mathbb Q}(X)=\rho^{\mathbb Q}(Y)$ for all $X,Y\in\X$.
One can refer to \cite{LW21} and \cite{LMWW22} for more details on the definition, properties and applications of tail risk measures.}

Next, we define the inf-convolution. The random variable $X\in \mathcal{X}$ represents a risk to be shared between $n$ agents. Then, define the set of \emph{allocations} of $X$ as
 \begin{equation*}
\mathbb{A}_n(X)=\left\{(X_1,\ldots,X_n)\in \mathcal{X}^n: \sum_{i=1}^nX_i=X\right\}. \label{eq:intro1}
\end{equation*}

Throughout this paper, we assume that the probability measures $\mathbb Q_1,\dots,\mathbb Q_n$  live on $(\Omega,\mathcal F)$ and, with the exception of Subsection \ref{sec:multiple_general}, the corresponding probability spaces are atomless. 
In this paper, we suppose the agents may have heterogeneous beliefs, represented by some probability measures $\mathbb Q_1,\dots, \mathbb Q_n$. 
The \emph{inf-convolution} of the risk measures $\rho_1,\dots,\rho_n$, under the respective probability measures $\mathbb Q_1,\dots,\mathbb Q_n$ is the mapping  $\dsquare_{i=1}^n \rho_i^{\mathbb Q_i}:\X\to[-\infty,\infty)$, defined as
\begin{equation*}\label{eq:inf-con}
 \dsquare_{i=1}^n \rho_i^{\mathbb Q_i} (X)  = \inf\left\{\sum_{i=1}^n\rho_i^{\mathbb Q_i}(X_i) : (X_1,\cdots,X_n)\in \mathbb A_n(X)  \right\}.
 \end{equation*}
 Note that $\rho_i$ is law-invariant under $\mathbb Q_i$, where $\mathbb Q_i$ represents the belief of the i-th agent.
In case of belief heterogeneity, the inf-convolution may not be law-invariant, i.e., if  $X\overset{\mathbb Q_i}{=} Y$ for all $i=1,\dots,n$, then $\dsquare_{i=1}^n \rho_i^{\mathbb Q_i} (X)\neq \dsquare_{i=1}^n \rho_i^{\mathbb Q_i} (Y)$ may hold; see more details in the Remark at the end of Section \ref{sec:conditional}. 

 

An $n$-tuple $(X_1,\cdots,X_n)\in \mathbb A_n(X)$ is called \emph{an optimal allocation} of $X$ for $(\rho_1,\dots,\rho_n)$ under $(\mathbb Q_1,\dots,\mathbb Q_n)$
if $\sum_{i=1}^{n}\rho_i^{\mathbb Q_i}(X_i)=\dsquare_{i=1}^n \rho_i^{\mathbb Q_i}(X)$.  The inf-convolution $\dsquare_{i=1}^n \rho_i^{\mathbb Q_i} (X)$ can be interpreted as  the smallest possible aggregate capital for the total risk $X$ in financial system, if $\rho_i^{\mathbb Q_i}(X_i)$ represents the capital charge for the i-th financial  institution to hold the risky position $X_i$. More economic interpretations on the inf-convolution can be found in e.g., \cite{D12}, \cite{R13} and \cite{ELW18}.

For the risk measures $(\rho_1,\dots,\rho_n)$ under $(\mathbb Q_1,\dots,\mathbb Q_n)$ and a total risk $X$, an allocation $(X_1,\cdots,X_n)\in \mathbb A_n(X)$ is \emph{Pareto-optimal} if for any other allocation $(Y_1,\dots,Y_n)\in \mathbb A_n(X)$, $\rho_i^{\mathbb Q_i}(Y_i)\leq \rho_i^{\mathbb Q_i}(X_i)$ for all $i=1,\dots,n$ implies $\rho_i^{\mathbb Q_i}(Y_i)=\rho_i^{\mathbb Q_i}(X_i)$ for all $i=1,\dots,n$.   For  finite-valued monetary risk measures, it is shown in \cite{ELW18} that an allocation  is optimal if and only if it is Pareto-optimal. Note that $\Lambda\VaR$ (defined in \eqref{lambdaL}) does not satisfy cash-additivity; see \cite{FMP14} and \cite{BP22}.  Hence it is not a monetary risk measure. However, one can still show that the optimal allocation of the inf-convolution of $\Lambda\VaR$ is Pareto-optimal.


Finally, we define the Lambda Value-at-Risk.  For $\Lambda:\R\to [0,1]$ and a probability measure $\mathbb Q$, the Lambda Value-at-Risk are given by
 \begin{align}\label{lambdaL}\Lambda\VaR^{\mathbb Q}(X)&=\inf\{x\in \R: F_X^{\mathbb Q}(x) \ge 1-\Lambda(x) \},\nonumber\\
 \Lambda\VaR^{+,\mathbb Q}(X)&=\sup\{x\in \R: F_X^{\mathbb Q}(x)<1-\Lambda(x) \},
 \end{align}
where $\inf\emptyset=\infty$ and $\sup\emptyset=-\infty$. Note that $\Lambda\VaR^{\mathbb Q}(X)=\Lambda\VaR^{+,\mathbb Q}(X)$ if $\Lambda$ is increasing; Otherwise, it may not be true; see Proposition 6 of \cite{BP22}.
We refer to \cite{BP22} for two other definitions of $\Lambda\VaR$.  If $\Lambda$ is a constant, then $\Lambda\VaR^{\mathbb Q}$ boils down to $\VaR^{\mathbb Q}$, i.e.,
$\VaR^{\mathbb Q}$ at level $p\in [0,1)$ is given by
 $$
\VaR_p^{\mathbb Q}(X) =\Lambda\VaR^{\mathbb Q}(X)=F_X^{\mathbb Q, -1}(1-p),~~~~X\in \mathcal  X,
 $$ for $\Lambda=p$. Moreover, by definition, $\VaR_1^{\mathbb Q}(X)=-\infty$.  Although $\VaR$ has been criticized from different angles, it has been widely applied  in practice for risk management due to its simplicity and possession of some nice properties; see \cite{MFE15} and the references therein for more detailed discussion on $\VaR$.  Compared to $\VaR$, $\Lambda\VaR$ is more flexible in the choice of $\Lambda$ functions; it satisfies cash subadditivity and quasi-star-shapedness for increasing $\Lambda$; and $\Lambda\VaR^+$ is able to capture the tail risk for decreasing $\Lambda$ functions; see e.g., \cite{FMP14}, \cite{HMP18} and \cite{HWWX24}. 


   We denote by $\mathcal H$ the collection of all right-continuous functions $\Lambda: \mathbb{R}\to [0,1]$. Hereafter, for any $\Lambda$, we denote $\lambda^-=\inf_{x\in\R}\Lambda(x)$ and $\lambda^+=\sup_{x\in\R}\Lambda(x)$. We say that a constant $\lambda$ is \emph{attainable} for $\Lambda$  if there exists $x\in\R$ such that $\Lambda(x)=\lambda$.


The interplay among $\mathbb Q_1, \dots, \mathbb Q_n$ plays an important role in the inf-convolution, which is our main concern in this paper. This can be seen from the following proposition. We say a mapping $\rho:\mathcal X\to\R$ is \emph{constant-preserving} if $\rho(c)=c$ for all $c\in\R$. Note that both $\Lambda\VaR^{\mathbb Q}$ and $\Lambda\VaR^{+,\mathbb Q}$ satisfy constant preservation.
\begin{proposition}\label{prop:general1} Suppose $\mathbb Q_1$ and $\mathbb Q_2$ are mutually singular, and  $\rho_2$  is constant-preserving. Then for $X\in\X$,
 $$\rho_1^{\mathbb Q_1}\dsquare \rho_2^{\mathbb Q_2}(X)=-\infty.$$
 \end{proposition}

The conclusion in Proposition \ref{prop:general1} implies that for any two monetary risk measures $\rho_1$ and $\rho_2$, $\rho_1^{\mathbb Q_1}\dsquare \rho_2^{\mathbb Q_2}(X)=-\infty$ if $\mathbb Q_1$ and $\mathbb Q_2$ are mutually singular. This means that if the beliefs of agents are too divergent, then each one of them can get unreasonably good deal by transferring risks between them. Then, the fact that the inf-convolution is $-\infty$ reflects that agents can always make their position better and hence there is no optimal allocation. In contrast, if $\mathbb Q_1=\mathbb Q_2$, then $\rho_1^{\mathbb Q_1}\dsquare \rho_2^{\mathbb Q_2}(X)>-\infty$ for many monetary risk measures $\rho_1$ and $\rho_2$; see e.g., \cite{ELW18} for quantile-based risk measures and \cite{FS08} for convex risk measures.  Note that the conclusion in Proposition \ref{prop:general1} can be easily extended to the case with $n$ agents for $n\geq 3$. 

\section{Inf-convolution of multiple $\Lambda\VaR$}\label{sec:multiple}
In this section, we investigate the inf-convolution of $n$ Lambda Value-at-Risk with heterogeneous beliefs represented by $\mathbb Q_1,\dots, \mathbb Q_n$, respectively. Suppose $\X\supseteq L^\infty$.  We find an expression for the inf-convolution and also the corresponding optimal allocation, covering the results of the inf-convolution of $\Lambda\VaR$ or $\VaR$ with homogeneous or heterogeneous beliefs in \cite{L24},  \cite{ELMW19} and \cite{ELW18}.  We first consider the general beliefs and then discuss three different cases about the relation of the beliefs: homogeneous beliefs, conditional beliefs, and general beliefs with two agents.
\subsection{General beliefs}\label{sec:multiple_general}
\tbl{In this subsection only, we assume that $\mathbb Q_1,\dots, \mathbb Q_n$ are general probability measures that are not required to be atomless.}
We first introduce the following notation.
Let $\Pi_n(\Omega)=\{(A_1,\dots, A_n): \cup_{i=i}^n A_i=\Omega,~ A_i\cap A_j=\emptyset, ~i\neq j\}$. For $\Lambda_1,\dots,\Lambda_n\in\mathcal H$ with  $0<\lambda_i^-\leq \lambda_i^+<1$, let $$\tbl{\Gamma_{\Lambda_1,\dots,\Lambda_n}(X)=\inf\left\{\sum_{j=1}^{n}y_j: \mathbb Q_i\left(\left\{X>\sum_{j=1}^{n}y_j\right\}\cap A_i\right)\leq \Lambda_i(y_i)~\text{for some}~(A_1,\dots, A_n)\in\Pi_n(\Omega)\right\}.}$$
Note that $\Gamma_{\Lambda_1,\dots,\Lambda_n}(X)\in [-\infty,\infty)$.
\begin{theorem}\label{Thmain} For $\Lambda_i\in\mathcal H$ with $0<\lambda_i^-\leq \lambda_i^+<1$, we have
\begin{align*}\label{eq:multiple}\dsquare_{i=1}^n\Lambda_i\VaR^{\mathbb Q_i}(X)=\Gamma_{\Lambda_1,\dots,\Lambda_n}(X).
\end{align*}
Moreover, if the minimizer of $\Gamma_{\Lambda_1,\dots,\Lambda_n}(X)$ exists denoted by $(y_1^*, \dots, y_n^*, A_1^*, \dots, A_n^*)$, then the optimal allocation is given by
\begin{equation}\label{Optimal}
X_i=\left(X-\sum_{j=1}^{n}y_j^*\right)\id_{A_i^*}+y_i^*,~i=1,\dots, n.
\end{equation}
\end{theorem}

The allocation \eqref{Optimal} can be understood as follows. The $i$-the agent is allocated a constant loss $y_i^*$ plus a contingent loss, which is the excess of the total risk over the aggregate cash allocation, restricted to the set $A_i^*$. The key feature of those sets is seen by the definition of $\Gamma_{\Lambda_1,\dots,\Lambda_n}(X)$: for each agent, their joint probability of (i) the set $A_i^*$ on which they suffer the contingent loss and (ii) the total loss exceeding the aggregate cash allocation, is restricted by their $\Lambda_i$ function.  

Note that Theorem \ref{Thmain} is  very general, covering Theorems 1, 2, 4 of \cite{L24} and Theorem 4 of \cite{ELMW19} as special cases. Compared with the results in \cite{L24}, our results in Theorem \ref{Thmain} offer semi-explicit formulas for the inf-convolution.  Later, we will specify some special relation of $\mathbb Q_1,\dots,\mathbb Q_n$ to obtain more explicit formulas for the inf-convolution. However, it is unclear what the inf-convolution of multiple $\Lambda\VaR$ under general heterogeneous beliefs is, leaving it as an open problem and deserving further study in future.  Moreover, Theorem \ref{Thmain} does not explicitly show when the inf-convolution is $-\infty$. The discussion on the finiteness of $\Gamma_{\Lambda_1,\dots,\Lambda_n}(X)$ is complicated as it heavily relies on the properties of Lambda functions and the relationship between $\mathbb Q_i$; see \cite{L24} and Proposition \ref{prop:general1} in Section \ref{sec:2}. This will be discussed in the next two propositions. Finally, note that if all $\Lambda_i$ are constants, then Theorem \ref{Thmain} boils down to Theorem 4 of \cite{ELMW19}, where the inf-convolution of $\VaR$ has been studied with belief heterogeneity.

Next, let us discuss the finiteness of $\Gamma_{\Lambda_1,\dots,\Lambda_n}(X)$. Let $x\wedge y=\min(x,y)$, $x\vee y=\max(x,y)$, $\bigwedge_{i=1}^n x_i=\min_{i=1}^n x_i$ and $\bigvee_{i=1}^n x_i=\max_{i=1}^n x_i$.

\begin{proposition}\label{prop:finite1} For $\Lambda_i\in\mathcal H$ with $0<\lambda_i^-\leq \lambda_i^+<1$, we have
\begin{enumerate}[(i)]
\item If $\bigvee_{i=1}^n\frac{\mathbb Q_i(A_i)}{\lambda_i^-}\leq 1$ for some $(A_1,\dots, A_n)\in\Pi_n(\Omega)$, then $\Gamma_{\Lambda_1,\dots,\Lambda_n}(X)=-\infty$;
\item If \tbl{$\inf_{(A_1,\dots, A_n)\in\Pi_n(\Omega)}\bigvee_{i=1}^n\frac{\mathbb Q_i(A_i)}{\lambda_i^+}>1$}, then $\Gamma_{\Lambda_1,\dots,\Lambda_n}(X)>-\infty$.
\end{enumerate}
\end{proposition}
Note that if all $\Lambda_i$ are constants and $\mathbb Q_1=\mathbb Q_2=\dots=\mathbb Q_n$, then Proposition \ref{prop:finite1} offers a sufficient and necessary condition for the finiteness of the inf-convolution. In general, due to the complexity of the $\Lambda_i$ functions and the heterogeneity of probability measures, those conditions are sufficient but not necessary.


As we can see from Proposition \ref{prop:finite1}, cases (i) and (ii) cannot cover all scenarios.
The discussion of other cases may require more information on the Lambda functions. We next consider all monotone Lambda functions in the same direction.
\begin{proposition}\label{prop:finite2}  Suppose all $\Lambda_i\in\mathcal H$ with $0<\lambda_i^-\leq \lambda_i^+<1$ are decreasing. Then we have the following conclusion.
\begin{enumerate}[(i)]
\item If \tbl{$\inf_{(A_1,\dots, A_n)\in\Pi_n(\Omega)}\bigvee_{i=1}^n\frac{\mathbb Q_i(A_i)}{\lambda_i^+}<1$}, then $\Gamma_{\Lambda_1,\dots,\Lambda_n}(X)=-\infty$;
\item If \tbl{$\inf_{(A_1,\dots, A_n)\in\Pi_n(\Omega)}\bigvee_{i=1}^n\frac{\mathbb Q_i(A_i)}{\lambda_i^+}>1$}, then $\Gamma_{\Lambda_1,\dots,\Lambda_n}(X)>-\infty$.
\end{enumerate}
Suppose all $\Lambda_i\in\mathcal H$ with $0<\lambda_i^-\leq \lambda_i^+<1$ are increasing. Then we have the following conclusion.
\begin{enumerate}[(i)]
\item If $\inf_{(A_1,\dots, A_n)\in\Pi_n(\Omega)}\bigwedge_{i=1}^n\left(\frac{\mathbb Q_i(A_i)}{\lambda_i^-}\vee\bigvee_{j\neq i}\frac{\mathbb Q_j(A_j)}{\lambda_j^+}\right)<1$, then $\Gamma_{\Lambda_1,\dots,\Lambda_n}(X)=-\infty$;
\item If $\inf_{(A_1,\dots, A_n)\in\Pi_n(\Omega)}\bigwedge_{i=1}^n\left(\frac{\mathbb Q_i(A_i)}{\lambda_i^-}\vee\bigvee_{j\neq i}\frac{\mathbb Q_j(A_j)}{\lambda_j^+}\right)>1$, then $\Gamma_{\Lambda_1,\dots,\Lambda_n}(X)>-\infty$.
\end{enumerate}
\end{proposition}

In this more refined result, we see that $\bigvee_{i=1}^n\frac{\mathbb Q_i(A_i)}{\lambda_i^+}$ plays a pivotal role when $\Lambda$ functions are decreasing.  We see that, in order to guarantee a finite inf-convolution for decreasing $\Lambda_i$, it is necessary for this quantity to exceed 1. For simplicity, assume that for all $i$, $\lambda_i^+$ is attained by $\Lambda_i$. Then, the condition in case (ii) of  Proposition \ref{prop:finite2} for deceasing $\Lambda_i$ is violated when one can find a partition of $\Omega$ such that for all agents we have $\mathbb Q_i(A_i) \leq \lambda_i^+$. This means that for each agent we can find a set $A_i$ on which the risk is suitably bounded by the largest acceptable probability to the agent of an adverse tail event. We can consider $\inf_{(A_1,\dots, A_n)\in\Pi_n(\Omega)} \bigvee_{i=1}^n\frac{\mathbb Q_i(A_i)}{\lambda_i^+}$ as a measure of belief homogeneity, which also reflects preferences. Consider two extreme cases. First, if sets $A_1,\dots,A_n$ exist for which we have $\mathbb Q_i(A_i)=0$ for all $i$, then we will have 
$\bigvee_{i=1}^n\frac{\mathbb Q_i(A_i)}{\lambda_i^+}=0.
$
The ability to find such sets is of course a case of extremely diverging beliefs. On the other hand, consider the case of belief homogeneity, $\mathbb Q_1=\dots=\mathbb Q_n=\mathbb P$. If $\mathbb P$ is atomless, then we will have
$
\inf_{(A_1,\dots, A_n)\in\Pi_n(\Omega)}\bigvee_{i=1}^n\frac{\mathbb Q_i(A_i)}{\lambda_i^+}>1
$
if and only if $\sum_{i=1}^n\lambda_i^+<1$, which reflects the impact of preferences. Clearly, when the values of $\sum_{i=1}^n\lambda_i^+<1$, indicating that agents accept lower total probabilities of adverse events, the inf-convolution is finite. Whereas, the higher risk tolerance of the tail events (i.e., $\sum_{i=1}^n \lambda_i^+>1$) enables the possibility of finding a trivially good deal for all agents, leading to an infinite inf-convolution.

When the functions are increasing, the picture is somewhat more complicated. Assuming attainability of all $\lambda_i^-,\lambda_i^+$, the condition of part (i) is satisfied if we can find a partition for which, for some $i$, we have that $\mathbb Q_i(A_i)<\lambda_i^-$ and $\mathbb Q_j(A_j)<\lambda_j^+,~j\neq i$. Hence, the conditions of Proposition \ref{prop:finite2}  compared to those in Proposition \ref{prop:finite1}, are weaker for part (i) and essentially weaker for part (ii), as the effect of the additional assumption of the $\Lambda$ functions.
\subsection{Homogeneous beliefs}
In this subsection, we focus on the case $\mathbb Q_1=\dots=\mathbb Q_n=\mathbb P$. We find more explicit expression for the inf-convolution.
For $x\in\R$, let $$\tbl{\Lambda^*(x)=\sup_{y_1+\dots+y_n=x}\left(1\wedge\sum_{i=1}^{n}\Lambda_i(y_i)\right)}.$$ We say $\Lambda^*$ is attainable if for each $x\in\R$, there exists $(y_1,\dots,y_n)\in\R^n$ satisfying $\sum_{i=1}^n y_i=x$ such that $1\wedge\sum_{i=1}^{n}\Lambda_i(y_i)=\Lambda^*(x)$. For homogeneous beliefs, $\Lambda^*(x)$ represents the largest aggregate risk tolerance of the agents to control the tail event $\{X>x\}$ in the definition of $\Gamma_{\Lambda_1,\dots,\Lambda_n}(X)$.
\begin{theorem}\label{prop:homogeneous} For $\Lambda_i\in\mathcal H$ with $0<\lambda_i^-\leq \lambda_i^+<1$, if $\Lambda^*$ is attainable, then we have
\begin{align*}\dsquare_{i=1}^n\Lambda_i\VaR^{\mathbb P}(X)=\Lambda^*\VaR^{\mathbb P}(X).
\end{align*}
Moreover, if additionally $x^*=\Lambda^*\VaR^{\mathbb P}(X)>-\infty$ and $\Lambda^*$ is right-continuous at $x^*$, then the optimal allocation is given by \eqref{Optimal} with  $(y_1^*, \dots, y_n^*)$  satisfying $1\wedge\sum_{i=1}^{n}\Lambda_i(y_i^*)=\Lambda^*(x^*)$ and $(A_1^*, \dots, A_n^*)$ satisfying $\mathbb P\left(\{X>x^*\}\cap A_i^*\right)\leq \Lambda_i(y_i^*)$.
\end{theorem}

Note that Theorem \ref{prop:homogeneous} shows that the optimal sets $(A_1^*, \dots, A_n^*)$ may not be unique. Under the assumption of  Theorem \ref{prop:homogeneous}, for $(y_1^*, \dots, y_n^*)$  satisfying $1\wedge\sum_{i=1}^{n}\Lambda_i(y_i^*)=\Lambda^*(x^*)$, optimal sets can be chosen as
\begin{align*}
 A_i^*&=\left\{1-\sum_{j=1}^{i} \Lambda_j(y_j^*)<U_X^{\mathbb P}\leq 1-\sum_{j=1}^{i-1} \Lambda_j(y_j^*)\right\}, ~i=1,\dots,n-1,\\
 A_n^*&=\left\{\tbl{\left(1-\sum_{j=1}^{n} \Lambda_j(y_j^*)\right)\vee 0}<U_X^{\mathbb P}\leq 1-\sum_{j=1}^{n-1} \Lambda_j(y_j^*)\right\},
\end{align*}
with the convention that $\sum_{j=1}^0 a_i=0$.
Moreover, the right-continuity of $\Lambda^*$ at $x^*$ can be removed if all $\Lambda_i$ are decreasing. Given the non-uniqueness of the allocation and the particular form of $A_i^*$, it is clear that agents' preferences do not impact which specific part of the (tail of) $X$ they are exposed to at the optimum. What is however impacted is the probability of the contingent events, specifically, for $\Lambda^*(x^*)<1$, we have $\mathbb P(A_i^*)= \Lambda_i(y_i^*),~i=1,\dots,n-1$ and 
$\mathbb P(A_n^*)= 1-\sum_{j=1}^{n-1} \Lambda_j(y_j^*).$

Note that in Theorem \ref{prop:homogeneous}, we do not require any monotonic properties of Lambda functions. The Lambda functions are very general with only one requirement on the attainability of $\Lambda^*$. Hence, it extends the results of Theorems 1, 2, 4 of \cite{L24}, who investigates the inf-convolution of multiple $\Lambda\VaR$ with monotone Lambda functions for homogeneous beliefs. In particular, Theorem \ref{prop:homogeneous} covers the case that some $\Lambda\VaR$ have increasing $\Lambda$ functions and others have decreasing $\Lambda$ functions, demonstrating that the agents may have relatively different risk appetites.

We can also remove the assumption on the attainability of $\Lambda^*$ in Theorem \ref{prop:homogeneous} to obtain an alternative expression.
For $n\geq 2$, let $$\tbl{\Lambda^{\mathbf{y}_{n-1}}(x)=\left(\Lambda_{n}(x-y_{n-1})+
\sum_{i=1}^{n-1}\Lambda_{i}(y_i-y_{i-1})\right)\wedge 1},$$ where   $\mathbf{y}_{n-1}=(y_1,\dots,y_{n-1})$ and $y_0=0$.
\begin{proposition}\label{prop:General}
For $\Lambda_i\in\mathcal H$ with $0<\lambda_i^-\leq \lambda_i^+<1$, we have
\begin{align*}\dsquare_{i=1}^n\Lambda_i\VaR^{\mathbb P}(X)=\inf_{\mathbf y_{n-1}\in\R^{n-1}}\Lambda^{\mathbf{y}_{n-1}}\VaR^{\mathbb P}(X).
\end{align*}
\end{proposition}
\tbl{Different from Theorem \ref{Thmain}, the assumption of measures being atomless cannot be removed in Theorem \ref{prop:homogeneous}. This can be seen from the following example.
\begin{example} For $m\geq 2$, suppose there exist distinct $w_1,w_2,\dots, w_m\in \Omega$ such that $\{w_1\},\dots,\{w_n\}\in \mathcal F$ and $\mathbb P({w_1})=\dots=\mathbb P({w_m})=1/m$. Define a random variable $X$ with $X(w_i)=i,~i=1,\dots, m$ and $X(w)=0$ if $w\notin \{w_1,\dots, w_m\}$. Clearly, $X$ is measurable under $(\Omega, \mathcal F)$. Moreover, suppose $\Lambda_1=\Lambda_2=1/(2m)$. It follows from Theorem \ref{Thmain} that 
\begin{align*}&~~\Lambda_1\VaR^{\mathbb P}\dsquare\Lambda_2\VaR^{\mathbb P}(X)\\
&=\inf\{y\in\R: \mathbb P(\{X>y\}\cap A))\vee \mathbb P(\{X>y\}\cap A^c))\leq 1/(2m)~\text{for some}~A\in \mathcal F\}=m.
\end{align*}
On the other hand, $$\Lambda^*\VaR^{\mathbb P}(X)=\VaR_{1/m}^{\mathbb P}(X)=m-1.$$
Hence, $\Lambda_1\VaR^{\mathbb P}\dsquare\Lambda_2\VaR^{\mathbb P}(X)\neq \Lambda^*\VaR^{\mathbb P}(X)$. This implies that the conclusion in Theorem \ref{prop:homogeneous} may not hold if the atomlessness assumption is removed.
\end{example}}
\subsection{Conditional beliefs}\label{sec:conditional}
In this subsection, we study further how the relationship among $\mathbb Q_1,\dots,\mathbb Q_n$ can affect the inf-convolution. In particular, we consider a case where agents operate with different information. To model this scenario, consider the case $\mathbb Q_i(\cdot)=\mathbb P(\cdot|B_i)$ for some $B_i\in\mathcal F$ with $\mathbb P(B_i)>0$. If $\mathbb P$ represents the physical probability, and $\mathbb Q_1,\dots,\mathbb Q_n$ are the beliefs of the agents, then those $\mathbb Q_i$ demonstrate that the agents' may have different assessments of the occurrence of events due to the asymmetric information or being subject to different regulating agencies with heterogeneous stressing events; see e.g., \cite{AS09}. 
For $B_1,\dots, B_n\in\mathcal F$ and $\Lambda_1,\dots,\Lambda_n\in\mathcal H$, if $\mathbb P(\cap_{i=1}^nB_i)>0$, let  $$\tbl{\Lambda^\diamond(x)=\sup_{y_1+\dots+y_n=x}\left(1\wedge\sum_{i=1}^{n}\frac{\mathbb P(B_i)\Lambda_i(y_i)}{\mathbb P(\cap_{j=1}^nB_j)}\right)}$$ for all $x\in\R$. Note that $\Lambda^\diamond(x)$ can be viewed as  the largest weighted aggregate risk tolerance of the agents to control the tail event $\{X>x\}$.

\begin{theorem}\label{prop:heterogeneous} Suppose $\mathbb Q_i(\cdot)=\mathbb P(\cdot|B_i)$ for some $B_i\subset\Omega$ with $\mathbb P(B_i)>0,~i=1,\dots,n$. For $\Lambda_i\in\mathcal H$ with $0<\lambda_i^-\leq \lambda_i^+<1$, if $\mathbb P(\cap_{i=1}^nB_i)=0$, then $$\dsquare_{i=1}^n\Lambda_i\VaR^{\mathbb Q_i}(X)=-\infty;$$
If $\mathbb P(\cap_{i=1}^nB_i)>0$ and $\Lambda^\diamond$ is attainable, then
\begin{align*}\dsquare_{i=1}^n\Lambda_i\VaR^{\mathbb Q_i}(X)=\Lambda^\diamond\VaR^{\mathbb Q}(X),
\end{align*}
where $\mathbb Q(\cdot)=\mathbb P(\cdot|\cap_{i=1}^nB_i)$.
Moreover, if additionally $x^*=\Lambda^\diamond\VaR^{\mathbb Q}(X)>-\infty$ and $\Lambda^\diamond$ is right-continuous at $x^*$, then the optimal allocation is given by \eqref{Optimal} with  $(y_1^*, \dots, y_n^*)$  satisfying $\sum_{i=1}^{n}\frac{\mathbb P(B_i)\Lambda_i(y_i^*)}{\mathbb P(\cap_{j=1}^n B_j)}=\Lambda^\diamond(x^*)$ and $(A_1^*, \dots, A_n^*)$ satisfying $\mathbb P\left(\{X>\sum_{i=1}^{n}y_i^*\}\cap\left(\cap_{i=1}^n B_i\right)\cap A_j^*\right)\leq \mathbb P(B_j)\Lambda_j(y_j^*)$.
\end{theorem}

Note that Theorem \ref{prop:heterogeneous} shows that the optimal sets $(A_1^*, \dots, A_n^*)$ may not be unique. Under the assumption of  Theorem \ref{prop:heterogeneous}, for $(y_1^*, \dots, y_n^*)$  satisfying $\sum_{i=1}^{n}\frac{\mathbb P(B_i)\Lambda_i(y_i^*)}{\mathbb P(\cap_{j=1}^n B_j)}=\Lambda^\diamond(x^*)$,  optimal sets can be chosen as 
\begin{align*}
  A_i^*&=\left\{1-\sum_{j=1}^{i} \frac{\mathbb P(B_j)\Lambda_j(y_j^*)}{\mathbb P(\cap_{k=1}^nB_k)}< U_X^{\mathbb Q}\leq 1-\sum_{j=1}^{i-1} \frac{\mathbb P(B_j)\Lambda_j(y_j^*)}{\mathbb P(\cap_{k=1}^nB_k)}\right\}, ~i=1,\dots,n-1,\\
   A_n^*&=\left\{\tbl{\left(1-\sum_{j=1}^{n} \frac{\mathbb P(B_j)\Lambda_j(y_j^*)}{\mathbb P(\cap_{k=1}^nB_k)}\right)\vee 0}< U_X^{\mathbb Q}\leq 1-\sum_{j=1}^{n-1} \frac{\mathbb P(B_j)\Lambda_j(y_j^*)}{\mathbb P(\cap_{k=1}^nB_k)}\right\}
\end{align*}
with the convention that $\sum_{j=1}^0 a_j=0$ and $\mathbb Q(\cdot)=\mathbb P(\cdot|\cap_{i=1}^nB_i)$.
Moreover, the right-continuity of $\Lambda^\diamond$ at $x^*$ can be removed if all $\Lambda_i$ are decreasing. 
Now, if $\Lambda^\diamond(x^*)<1$, the probability of the optimal sets, under the probability $\mathbb Q$ representing evaluations with respect to the information set that all agents agree on,  are given by
$$
\mathbb Q(A_i^*)=\frac{\mathbb P(B_i)\Lambda_i(y_i^*)}{\mathbb P(\cap_{k=1}^nB_k)},\quad i=1,\dots,n-1, ~\mathbb Q(A_n^*)=1-\sum_{j=1}^{n-1} \frac{\mathbb P(B_j)\Lambda_j(y_j^*)}{\mathbb P(\cap_{k=1}^nB_k)},
$$
or, equivalently,
$$
\mathbb P\left((\cap_{k=1}^n B_k) \cap A_i^*\right) = \mathbb P(B_i)\Lambda_i(y_i^*),~i=1,\dots,n-1,~\mathbb P(A_n^*)=\mathbb P(\cap_{k=1}^nB_k)-\sum_{j=1}^{n-1} \mathbb P(B_j)\Lambda_j(y_j^*).
$$

In this special case, the interplay among $\mathbb Q_1,\dots,\mathbb Q_n$ is characterized by the relationship between the sets $B_1,\dots, B_n$. Following the conclusion in Proposition \ref{prop:heterogeneous}, those sets $B_1,\dots, B_n$ appear in the expressions of the inf-convolution and the optimal allocation, suggesting the influence of the relation of $\mathbb Q_1,\dots,\mathbb Q_n$ on the inf-convolution. In particular,  if the agents have a strong disagreement on the interested events such that $\mathbb P(\cap_{i=1}^nB_i)=0$, it leads to a trivial case. Note that the crucial difference to the homogeneous case is that here, in characterising the allocations, we use the set $ \left(\cap_{k=1}^n B_k\right)\cap A_i^*$ which includes both the set  $A_i^*$ on which the contingent loss is paid and the conditioning set (information) on which all agents can agree.
If all Lambda functions in Proposition \ref{prop:heterogeneous} are assumed to be constants, then we immediately derive the explicit formulas for the inf-convolution of multiple $\VaR$ under conditional beliefs.
\begin{corollary}\label{cor:var} Suppose $\mathbb Q_i(\cdot)=\mathbb P(\cdot|B_i)$ for some $B_i\subset\Omega$ with $\mathbb P(B_i)>0,~i=1,\dots,n$. For $\alpha_i\in (0,1)$, if $\mathbb P(\cap_{i=1}^nB_i)=0$, then $$\dsquare_{i=1}^n\VaR_{\alpha_i}^{\mathbb Q_i}(X)=-\infty;$$
If $\mathbb P(\cap_{i=1}^nB_i)>0$, then
\begin{align*}\dsquare_{i=1}^n\VaR_{\alpha_i}^{\mathbb Q_i}(X)=\VaR_{1\wedge\sum_{i=1}^{n}\frac{\alpha_i\mathbb P(B_i)}{\mathbb P(\cap_{j=1}^nB_j)}}^{\mathbb Q}(X),
\end{align*}
where $\mathbb Q(\cdot)=\mathbb P(\cdot|\cap_{i=1}^nB_i)$.
\end{corollary}
The results in Corollary \ref{cor:var} extend the elegant formula for the inf-convolution of $\VaR$ under homogeneous beliefs to the case with conditional beliefs; see Corollary 2 of \cite{ELW18}. The simplicity of the expressions for $\VaR$ allow some additional insight. We can see that the trivial outcome $\dsquare_{i=1}^n\VaR_{\alpha_i}^{\mathbb Q_i}(X)=-\infty$ is not only achieved in the case of full disagreement between agents $\mathbb P(\cap_{i=1}^nB_i)=0$. It also occurs under weaker disagreement, when  $\sum_{i=1}^{n}\frac{\alpha_i\mathbb P(B_i)}{\mathbb P(\cap_{j=1}^nB_j)}\geq 1$. This inequality may be satisfied, when $\mathbb P(\cap_{j=1}^nB_j)$ is sufficiently small, indicating high -- but partial -- disagreement in conditional beliefs, and the levels $\alpha_i$ are sufficiently large, indicating a high level of risk tolerance.

It is worth mentioning that the results in Proposition \ref{prop:heterogeneous} can be explained from the perspective of Co-risk measures.  Recently, Co-risk measures are introduced in the context of systemic risk measurement, with reference to conditional $\VaR$ \citep{AB16,GE13} and conditional $\ES$ \citep{MS14} .  Let $B\in\mathcal F$ satisfying $\mathbb P(B)>0$. For $\alpha\in (0,1)$,  the conditional $\VaR$ $(\mathrm{Co\VaR})$ and conditional $\ES$ $(\mathrm{Co\ES})$ are defined respectively as
\begin{align*}
\VaR_{\alpha}^{\mathbb P}(X|B)&=\inf\{x: \mathbb P(X\leq x|B)\geq 1-\alpha\},~
 \ES_{\alpha}^{\mathbb P}(X|B)=\frac{1}{\alpha}\int_{0}^{\alpha}\VaR_{t}^{\mathbb P}(X|B)\d t.
\end{align*}
Note that if $B=\Omega$, conditional $\VaR$ and $\ES$ boil down to the classical $\VaR$ and $\ES$, i.e., $\VaR_{\alpha}^{\mathbb P}(X)=\VaR_{\alpha}^{\mathbb P}(X|\Omega)$ and $\ES_{\alpha}^{\mathbb P}(X)=\ES_{\alpha}^{\mathbb P}(X|\Omega)$.
Analogously, the conditional $\Lambda\VaR$ ($\mathrm{Co\Lambda\VaR}$) can be defined  as $$\Lambda\VaR^{\mathbb P}(X|B)=\inf\{x: \mathbb P(X\leq x|B)\geq \Lambda(x)\}.$$
Note that  $\Lambda_i\VaR^{\mathbb Q_i}(X)=\Lambda_i\VaR^{\mathbb P}(X|B_i)$ if $\mathbb Q_i(\cdot)=\mathbb P(\cdot|B_i)$.
Hence, the results in Theorem \ref{prop:heterogeneous} can be interpreted as the risk sharing using multiple $\mathrm{Co\Lambda\VaR}$ to represent the preference of the agents. In the regulatory context, those $B_i$ could be stressing test events given by different supervising agencies.  

\begin{remark*}
The law-invariance of the inf-convolution for some risk functionals was discussed in \cite{LWW20}, showing that the law-invariance of the inf-convolution under homogeneous beliefs is guaranteed by some mild condition on the risk functionals if they are law-invariant. Proposition \ref{prop:General} shows that the inf-convolution of $\Lambda\VaR$ under homogeneous beliefs is law-invariant. However, if the beliefs are heterogeneous, the conclusion may not hold, which can be seen from Corollary \ref{cor:var}.  Suppose $\mathbb Q_i(\cdot)=\mathbb P(\cdot|B_i)$ for some $B_i\subset\Omega$ with $\mathbb P(\cap_{i=1}^n B_i)>0$ and $\alpha_i\in (0,1)$ satisfying $\sum_{i=1}^{n}\frac{\alpha_i\mathbb P(B_i)}{\mathbb P(\cap_{j=1}^nB_j)}<1$. By Corollary  \ref{cor:var}, we have \begin{align*}\dsquare_{i=1}^n\VaR_{\alpha_i}^{\mathbb Q_i}(X)=\VaR_{\sum_{i=1}^{n}\frac{\alpha_i\mathbb P(B_i)}{\mathbb P(\cap_{j=1}^nB_j)}}^{\mathbb Q}(X),
\end{align*}
where $\mathbb Q(\cdot)=\mathbb P(\cdot|\cap_{i=1}^nB_i)$.  One can easily check that  $X\overset{\mathbb Q_i}{=} Y$ for all $i=1,\dots,n$ does not imply  $X\overset{\mathbb Q}{=} Y$. Hence, $\dsquare_{i=1}^n\VaR_{\alpha_i}^{\mathbb Q_i}(X)\neq \dsquare_{i=1}^n\VaR_{\alpha_i}^{\mathbb Q_i}(Y)$ may hold.
\end{remark*}

 \subsection{General beliefs with $n=2$}
  Next, we consider the case $n=2$. \tbl{By Theorem A.17 of \cite{FS16}, for probability measures $\mathbb Q_1$ and $\mathbb Q_2$, we have 
  \begin{align}\label{Eq:AC}\mathbb Q_2(B)=\mathbb E^{\mathbb Q_1}(\eta\id_B)+\mathbb Q_2(B\cap N)~\text{for all}~B\in \mathcal F,
  \end{align}
  where $N\in\mathcal F$ satisfying $\mathbb Q_1(N)=0$ and $\eta=\frac{d\mathbb Q_2}{d\mathbb Q_1}\id_{N^c}$.  If $\mathbb Q_2(N)=0$, then $\mathbb Q_2<<\mathbb Q_1$ on $\Omega$; if $\mathbb Q_2(N)>0$, then  $\mathbb Q_2<<\mathbb Q_1$ on $N^c$ but not on $N$. In this subsection, the relation of $\mathbb Q_1$ and $\mathbb Q_2$ is represented by \eqref{Eq:AC}.}
\begin{theorem}\label{prop:continuous}  For $\Lambda_i\in\mathcal H$ with $0<\lambda_i^-\leq \lambda_i^+<1$, we have
\begin{align*}\Lambda_1\VaR^{\mathbb Q_1}\dsquare \Lambda_2\VaR^{\mathbb Q_2}(X)=\inf\left\{x\in\R: g_{x}((1-F_X^{\mathbb Q_1}(x)-\Lambda_1(y))_+)\leq \Lambda_2(x-y),~\text{for some}~y\in\R\right\},
\end{align*}
where $$g_{x}(t)=\mathbb E^{\mathbb Q_1}(\eta\id_{\{X>x\}\cap \{U_\eta^{\mathbb Q_1}\leq x_t\}}),~t\in [0,1-F_X^{\mathbb Q_1}(x)],$$ with $x_t$ satisfying $\mathbb Q_1(\{X>x\}\cap \{U_\eta^{\mathbb Q_1}\leq x_t\})=t$. 
Moreover, if $x^*=\Lambda_1\VaR^{\mathbb Q_1}\dsquare \Lambda_2\VaR^{\mathbb Q_2}(X)>-\infty$ and $g_{x^*}((1-F_X^{\mathbb Q_1}(x^*)-\Lambda_1(y^*))_+)\leq \Lambda_2(x^*-y^*)$ for some $y^*\in\R$, then the optimal allocation is given by
\begin{equation}\label{OptimalAb}
X_1=\left(X-x^*\right)\id_{\Omega\setminus A^*}+y^*,~X_2=\left(X-x^*\right)\id_{A^*}+x^*-y^*,
\end{equation}
where \tbl{$A^*=\{X>x^*\}\cap\{U_\eta^{\mathbb Q_1}\leq z^*\}\cap N^c$} with $z^*$ satisfying $\mathbb Q_1(\{X>x^*\}\cap \{U_\eta^{\mathbb Q_1}\leq z^*\})=(1-F_X^{\mathbb Q_1}(x^*)-\Lambda_1(y^*))_+$.
\end{theorem}
\tbl{An immediate conclusion from Theorem \ref{prop:continuous} is the following result, showing that the inf-convolution is $-\infty$ if $\mathbb Q_2(N)$ is large enough.
\begin{corollary}\label{Cor:new} Under the assumption of Theorem \ref{prop:continuous}, if $\mathbb Q_2(N)\geq 1-\lambda_2^-$, then 
    $$\Lambda_1\VaR^{\mathbb Q_1}\dsquare \Lambda_2\VaR^{\mathbb Q_2}(X)=-\infty.$$
\end{corollary}}

\tbl{Note that the idea used in the proof of Theorem \ref{prop:continuous} is not valid for the case $n\geq 3$. For the case $n\geq 3$, it may require a new method to simplify the main result obtained in Theorem \ref{Thmain}.}

In Theorem \ref{prop:continuous}, we do not obtain explicit formulas as the previous propositions. This is due to the complexity of the relation between $\mathbb Q_1$ and $\mathbb Q_2$, which can be observed from the expression of $g_x$ depending on the joint distribution of $(X,\eta)$ under $\mathbb Q_1$. Let us next see some examples of $g_x$ and the corresponding $A^*$ in \eqref{OptimalAb}:
\begin{enumerate}[(i)]
\item If $X$ and $\eta$ are comonotonic, then $g_x(t)=\int_{F_X^{\mathbb Q_1}(x)}^{F_X^{\mathbb Q_1}(x)+t} F_{\eta}^{\mathbb Q_1, -1}(s)\d s,~t\in [0,1-F_X^{\mathbb Q_1}(x)]$ and $A^*=\{X>x^*\}\cap\{U_\eta^{\mathbb Q_1}\leq (1-\Lambda_1(y^*))\vee F_X^{\mathbb Q_1}(x^*)\}\cap N^c$; 
\item If  $X$ and $\eta$ are countermonotonic, then 
$g_x(t)=\int_0^t F_{\eta}^{\mathbb Q_1, -1}(s)\d s,~t\in [0,1-F_X^{\mathbb Q_1}(x)]$ and $A^*=\{X>x^*\}\cap\{U_\eta^{\mathbb Q_1}\leq (1-F_X^{\mathbb Q_1}(x^*)-\Lambda_1(y^*))_+\}\cap N^c$;
\item If $X$ and $\eta$ are independent under $\mathbb Q_1$, then $g_x(t)=(1-F_X^{\mathbb Q_1}(x))\int_{0}^{t/(1-F_X^{\mathbb Q_1}(x))} F_{\eta}^{\mathbb Q_1, -1}(s)\d s,~t\in [0,1-F_X^{\mathbb Q_1}(x)]$ with the convention that $0/0=0$ and $$A^*=\{X>x^*\}\cap\left\{ U_\eta^{\mathbb Q_1}\leq \frac{(1-F_X^{\mathbb Q_1}(x^*)-\Lambda_1(y^*))_+}{1-F_X^{\mathbb Q_1}(x^*)}\right\}\cap N^c$$ with $U_\eta^{\mathbb Q_1}$ being independent of $X$ under $\mathbb Q_1$.
\end{enumerate}
Note that the optimal set $A^*$ in (i) and (ii) have very different structures. To demonstrate this, suppose both $X$ and $\eta$ are continuous random variables under $\mathbb Q_1$ and $N=\emptyset$.  In (i), $A^*$ can be rewritten as $A^*=\{F_X^{\mathbb Q_1}(x^*)<U_X^{\mathbb Q_1}<(1-\Lambda_1(y^*))\vee F_X^{\mathbb Q_1}(x^*)\}$. Hence, the contingent loss of agent 2 takes place on a set that does not include the full right tail of $X$. This is understandable because the comonotonicity of $X$ and $\eta$ means that agent 2 considers the distribution of aggregate loss more risky in first-order stochastic dominance, {$F_X^{\mathbb Q_2}\succeq_{\mathrm{st}} F_X^{\mathbb Q_1}$} (see e.g. \cite{PMT19}). In the alternative case (ii) of comonotonicity, we can write $A^*=\{U_X^{\mathbb Q_1}>F_X^{\mathbb Q_1}(x^*)+\Lambda_1(y^*)\}$, such that agent 2 is exposed to the full tail of $X$, consistent with {$F_X^{\mathbb Q_1}\succeq_{\mathrm{st}} F_X^{\mathbb Q_2}$}. Hence, by \eqref{OptimalAb}, we see that if $X$ and $\eta$ are comonotonic, a lower contingent loss is allocated to the second
agent; if $X$ and $\eta$ are countermonotonic, a higher contingent loss is allocated to the second agent.


Note that Theorem \ref{prop:continuous} can be simplified when the Lambda functions are constants.
\begin{corollary}\label{cor:varas}
 For $\alpha_1,\alpha_2\in (0,1)$, we have 
\begin{align*}\VaR_{\alpha_1}^{\mathbb Q_1}\dsquare \VaR_{\alpha_2}^{\mathbb Q_2}(X)
=\inf\left\{x\in\R: g_{x}((1-F_X^{\mathbb Q_1}(x)-\alpha_1)_+)\leq \alpha_2\right\},
\end{align*}
where $g_{x}(t)=\mathbb E^{\mathbb Q_1}(\eta\id_{\{X>x, U_\eta^{\mathbb Q_1}\leq x_t\}}),~t\in [0,1-F_X^{\mathbb Q_1}(x)]$, with $x_t$ satisfying $\mathbb Q_1(\{X>x\}\cap\{U_\eta^{\mathbb Q_1}\leq x_t\})=t$.
\end{corollary}

 Under the assumption of Corollary \ref{cor:varas}, we have the following results.
 \begin{corollary}
\begin{enumerate}[(i)]
\item  If $X$ and $\eta$ are comonotonic, then
\begin{align*}\VaR_{\alpha_1}^{\mathbb Q_1}\dsquare \VaR_{\alpha_2}^{\mathbb Q_2}(X)
=\inf\left\{x\in\R: \int_{F_X^{\mathbb Q_1}(x)}^{F_X^{\mathbb Q_1}(x)\vee (1-\alpha_1)}F_{\eta}^{\mathbb Q_1, -1}(s)\d s\leq \alpha_2\right\};
\end{align*}
\item  If $X$ and $\eta$ are countermonotonic, then
\begin{align*}\VaR_{\alpha_1}^{\mathbb Q_1}\dsquare \VaR_{\alpha_2}^{\mathbb Q_2}(X)
= \inf\left\{x\in\R: \int_{0}^{(1-F_X^{\mathbb Q_1}(x)-\alpha_1)_+}F_{\eta}^{\mathbb Q_1, -1}(s)\d s\leq \alpha_2\right\};
\end{align*}
\item If $X$ and $\eta$ are independent under $\mathbb Q_1$, then
\begin{align*}\VaR_{\alpha_1}^{\mathbb Q_1}\dsquare \VaR_{\alpha_2}^{\mathbb Q_2}(X)
=\inf\left\{x\in\R: (1-F_X^{\mathbb Q_1}(x))\int_{0}^{(1-F_X^{\mathbb Q_1}(x)-\alpha_1)_+/(1-F_X^{\mathbb Q_1}(x))}F_{\eta}^{\mathbb Q_1, -1}(s)\d s\leq \alpha_2\right\}
\end{align*}
with the convention that $0/0=0$.
\end{enumerate}
\end{corollary}

\section{Inf-convolution of $\Lambda\VaR$ and one monotone risk measure with heterogeneous beliefs}\label{sec:5}
In this section, we investigate the risk sharing problem between two agents, considering the inf-convolution of $\Lambda\VaR$ and a monotone risk measure $\rho$ under the probability measures $\mathbb Q_1$ and $\mathbb Q_2$, where $\mathbb Q_2$ may be different from $\mathbb Q_1$. Note that $\rho$ here is very general including many commonly used risk measures as special cases. \tbl{In our results below, we consider both the cases $\X=L^\infty$ and unbounded $\X$, each with different assumptions on $\rho$.}

\begin{theorem}\label{Thone}
Suppose $\Lambda\in\mathcal H$ with $0<\lambda^-\leq \lambda^+<1$ and $\rho$ is monotone. \tbl{If either (i) $\X=L^\infty$ and $\rho$ is law-invariant under $\mathbb Q_2$ or (ii) $\X$ is unbounded and $\rho$ is an $\epsilon$-tail risk measure under $\mathbb Q_2$ for some $\epsilon \in (0,1)$},  then we have
\begin{equation}\label{Minimizer1}\Lambda\VaR^{\mathbb Q_1}\dsquare \rho^{\mathbb Q_2}(X)=\inf_{x\in\R}\inf_{y\in\R}\inf_{\mathbb Q_1(B)=1-\Lambda(x)}\left\{x+\rho^{\mathbb Q_2}((X-x)\id_{B}+y\id_{B^c})\right\}.
\end{equation}
Moreover, the existence of the optimal allocation for the inf-convolution is equivalent to the existence of the minimizer of \eqref{Minimizer1}. If $(x^*,y^*, B^*)$ is the minimizer, then one optimal allocation is given by
\begin{align}\label{Minimizer20}
X_1^*=(X-x^*-y^*)\id_{(B^*)^c}+x^*,~ X_2^*=(X-x^*-y^*)\id_{B^*}+y^*.
\end{align}
\end{theorem}
The optimal allocation in \eqref{Minimizer20} has the same form as that of Theorem \ref{Thmain}: each agent is allocated a constant risk and a contingent risk, which is the excess of the total risk over the aggregate cash allocation restricted to some contingent event. 
Note that the expression of the inf-convolution in Theorem \ref{Thone} involves an optimization problem with three parameters $(x,y,B)$, where $B$ is a measurable set contained in $\mathcal F$. The set $B$ makes this optimization problem complicated. The optimal allocation in \eqref{Minimizer20} may not be pairwise countermonotonic.

Our conclusion in Theorem \ref{Thone} extends the results of \cite{ELMW19} for $n=2$, where the inf-convolution of $\VaR$ or $\ES$ with heterogeneous beliefs is considered. If $\mathbb Q_1=\mathbb Q_2$, then  Theorem \ref{Thone} coincides with the results in Theorem 3 of \cite{L24}.
We notice that for the scenario that  $\mathbb{Q}_1=\mathbb{Q}_2$ and $\Lambda$ is a constant, Theorem \ref{Thone}  extends the results in Theorem 2 of \cite{LMWW22} ($\rho$ is there considered to be  a monetary tail risk measure) and
 Theorem 5.3 of \cite{WW20A} ($\rho$ is there a distortion risk measure). The result of Theorem \ref{Thone} is still new even if $\Lambda\VaR$ is reduced to $\VaR$.
Let us now consider  the case $\mathbb Q_i(\cdot)=\mathbb P(\cdot|B_i)$ for some $B_i\in\mathcal F$ with $\mathbb P(B_i)>0$, where $\mathbb P$ is atomless. Under this setup, the contingent set $B$ in \eqref{Minimizer1} is given explicitly.

\begin{proposition}\label{Pro:conditional}
 Suppose $\mathbb Q_i(\cdot)=\mathbb P(\cdot|B_i)$ for some $B_i\subset\Omega$ with $\mathbb P(B_i)>0,~i=1, 2$.  Moreover, suppose $\Lambda\in\mathcal H$ with $0<\lambda^-\leq \lambda^+<1$ and $\rho$ is monotone.  \tbl{If either (i) $\X=L^\infty$ and $\rho$ is law-invariant under $\mathbb Q_2$ or (ii) $\X$ is unbounded and $\rho$ is an $\epsilon$-tail risk measure under $\mathbb Q_2$ for some $\epsilon \in (0,1)$}, then we have
\begin{equation}\label{Minimizer11}\Lambda\VaR^{\mathbb Q_1}\dsquare \rho^{\mathbb Q_2}(X)=\inf_{x\in\R}\inf_{y\in\R}\left\{x+\rho^{\mathbb Q_2}((X-x)\id_{A_x}+y\id_{A_x^c})\right\},
\end{equation}
where $A_x=\emptyset$ if $\mathbb P(B_1\cap B_2^c)\geq (1-\Lambda(x))\mathbb P(B_1)$ and $A_x=(B_1\cap B_2^c)\cup(B_1\cap B_2\cap\{U_X^{\mathbb P}<\alpha_x\})$ if $\mathbb P(B_1\cap B_2^c)<(1-\Lambda(x))\mathbb P(B_1)$ with $\alpha_x$ satisfying $\mathbb P(B_1\cap B_2\cap\{U_X^{\mathbb P}<\alpha_x\})=(1-\Lambda(x))\mathbb P(B_1)-\mathbb P(B_1\cap B_2^c)$.
Moreover, the existence of the optimal allocation of the inf-convolution is equivalent to the existence of the minimizer of \eqref{Minimizer11}. If $(x^*,y^*)$ is the minimizer, then one optimal allocation is given by
\begin{align}\label{Minimizer30}
X_1^*=(X-x^*-y^*)\id_{A_{x^*}^c}+x^*,~ X_2^*=(X-x^*-y^*)\id_{A_{x^*}}+y^*.
\end{align}
\end{proposition}
Note that in the optimal allocation given by \eqref{Minimizer30}, the contingent event $A_{x^*}$ is a fixed event, making the optimal allocation more concrete than that of Theorem \ref{Thone}. 
We next interpret the optimal allocation given by \eqref{Minimizer30} by focusing on the case  $A_{x^*}=\emptyset$.  Note that the condition for $A_{x^*}=\emptyset$ can be reformulated as the condition:
\begin{align*}
    \mathbb P(B_1\cap B_2^c)\geq (1-\Lambda(x^*))\mathbb P(B_1) \Leftrightarrow \mathbb Q_1(B_2)\leq \Lambda (x^*).
\end{align*}
Clearly, $A_{x^*}=\emptyset$ means that all the risk stays with agent 1. The condition $\mathbb Q_1(B_2)\leq \Lambda (x^*)$ is satisfied when
$\mathbb Q_1(B_2)$ is small (this means that agent 1 dismisses the view of agent 2, so we have more information asymmetry); or when $\Lambda (x^*)$ is large (this means that agent 1 has a higher risk tolerance.)

 Let $\mathcal G=\{g:[0,1]\to[0,1]|~ g~\text{is increasing and left-continuous satisfying}~g(0)=0~\text{and}~g(1)=1\}$.  For $g\in\mathcal H$, the distortion risk measure $\rho_g$ under $\mathbb Q$ is defined as
\begin{align*}
 \rho_g^{\mathbb Q}(X)&=\int_{0}^{1}\VaR_{t}^{\mathbb Q}(X)\d g(t).
\end{align*}
 Distortion risk measures form a popular class of risk measures applied in insurance pricing, performance evaluation, decision theory and many other topics; see e.g., \cite{FS16}, \cite{Wang96}, \cite{Wang00}, \cite{CM09} and \cite{Y87}. Note that if $\mathbb Q(\cdot)=\mathbb P(\cdot|B)$ for some $B\in\mathcal F$ satisfying $\mathbb P(B)>0$, then $\rho_g^{\mathbb Q}(X)$ corresponds to the   Co-distortion risk
measures introduced in  \cite{DLZ22}. 

 In the following corollary, let $\mathbb Q_i(\cdot)=\mathbb P(\cdot|B_i)$ for some $B_i\subset\Omega$ with $\mathbb P(B_1\cap B_2)>0$, indicating that beliefs are not totally incompatible, and $\mathbb Q(\cdot)=\mathbb P(\cdot|B_1\cap B_2) $, which in a sense represents the consensus view of the two agents. Applying Proposition \ref{Pro:conditional}, we obtain the inf-convolution of $\mathrm{\Lambda\VaR}$ and $\rho_g/\Lambda_1\VaR^{+}$ under heterogeneous beliefs, respectively. In order to simplify the expression, we suppose $\lambda^+$ is attainable for $\Lambda$, i.e., there is $x\in\R$ such that $\Lambda(x)=\lambda^+$.

\begin{corollary}\label{Cor2} Suppose $\Lambda\in\mathcal H$ with $0<\lambda^-\leq \lambda^+<1$ and $\lambda^+$ is attainable.  Then the following conclusion holds with $\beta=\frac{(1-\lambda^+)\mathbb P(B_1)-\mathbb P(B_1\cap B_2^c)}{\mathbb P(B_2)}$.
\begin{enumerate}[(i)]
\item For $g\in\mathcal G$ \tbl{satisfying $g(\epsilon)=1$ for some $\epsilon\in (0,1)$}, if $\mathbb Q_1(B_2)\leq\lambda^+$, then $\Lambda\VaR^{\mathbb Q_1}\dsquare \rho_g^{\mathbb Q_2}(X)=-\infty$; if $\mathbb Q_1(B_2)>\lambda^+$ then
$$\Lambda\VaR^{\mathbb Q_1}\dsquare \rho_g^{\mathbb Q_2}(X)=\left\{\begin{array}{cc}
\int_{[0,\beta)} \VaR_{1-\frac{\mathbb P(B_2)(\beta-t)}{\mathbb P(B_1\cap B_2)}}^{\mathbb Q}(X)\d g(t),&g(\beta)=1 \\
-\infty,& g(\beta)<1
\end{array}\right.;$$
\item  For $\Lambda_1\in\mathcal H$ with $0<\lambda_1^-\leq \lambda_1^+<1$, if $\mathbb Q_1(B_2)\leq \lambda^+$, then $\Lambda\VaR^{\mathbb Q_1}\dsquare \Lambda_1\VaR^{+,\mathbb Q_2}(X)=-\infty$; if $\mathbb Q_1(B_2)>\lambda^+$, then  $$\Lambda\VaR^{\mathbb Q_1}\dsquare \Lambda_1\VaR^{+,\mathbb Q_2}(X)=\inf_{x\in\R}\overline{\Lambda}^x\VaR^{+,\mathbb Q}(X),$$
where $\overline{\Lambda}^x(z)=\frac{\mathbb P(B_1)\Lambda(x)+\mathbb P(B_2)\Lambda_1(z-x)}{\mathbb P(B_1\cap B_2)}\wedge 1$.
\end{enumerate}
\end{corollary}
Note that if we let  $g(t)=(t/\alpha)\wedge 1$ for $\alpha\in (0,1)$, then the distortion risk measure is the Expected Shortfall, i.e., $\rho_g=\ES_\alpha$. In that case, part (i) of Corollary \ref{Cor2} simplifies  to stating that, if $\mathbb Q_1(B_2)>\lambda^+$, then $$\Lambda\VaR^{\mathbb Q_1}\dsquare \ES_{\alpha}^{\mathbb Q_2}(X)=\left\{\begin{array}{cc}
\frac{1}{\alpha}\int_0^\alpha \VaR_{1-\frac{\mathbb P(B_2)(\beta-t)}{\mathbb P(B_1\cap B_2)}}^{\mathbb Q}(X)\d t,&\alpha\leq\beta \\
-\infty,& \alpha>\beta
\end{array}\right..$$
For all three cases in Corollary \ref{Cor2} the condition $\mathbb Q_1(B_2)\leq \lambda^+$, indicating high levels of belief heterogeneity / high level of risk tolerance for agent 1, gives a sufficient condition for the inf-convolution to be infinity. For part (i), we also get an infinite allocation as long as the quantity $\beta$ is small enough, such that $g(\beta)<1$. Note that in the special case where $\rho_g$ is Expected Shortfall at level $\alpha$, this means $\beta <\alpha$. To interpret small values of $\beta$ we can reformulate this quantity as follows:
$$
\beta=\frac{\mathbb P(B_1)}{\mathbb P(B_2)}\left(\mathbb Q_1(B_2)-\lambda^+ \right).
$$
It is seen that $\beta$ becomes small if 
\begin{itemize}
    \item $\lambda^+$ is high, such that agent 1 has a high maximal risk tolerance;
    \item $\mathbb Q_1(B_2)$ is low, so we have more belief heterogeneity;
    \item $\frac{\mathbb P(B_1)}{\mathbb P(B_2)}$ is low, since we are in the case that $\mathbb Q_1(B_2)>\lambda^+$. This fraction can be understood, in some sense, as a comparison of the subjective informativeness of the conditioning done by the two agents. If a low probability event is observed, one could claim that more has been learned about the underlying risk. Hence, if the ratio is small, agent 1 has subjectively learned more than agent 2 and is, in that specific sense, more confident. 
\end{itemize}
These interpretations tell us that we have an infinite allocation when there is belief heterogeneity and agent 1 is more risk tolerant as well as considering that the event they believe they observed is highly informative.

We mention that the results in Corollary \ref{Cor2} extend the results in the literature in the sense that $\Lambda$ is not a constant and the second risk measure is  not law-invariant under $\mathbb Q_1$. For instance, (i) of Corollary \ref{Cor2} exactly extends  Theorem 5.3 of \cite{WW20A} ($\rho$ is a distortion risk measure which is law-invariant under $\mathbb Q_1$) in these two directions. It is worth mentioning that (ii) of Corollary \ref{Cor2} is a new result even for the case $\mathbb Q_1=\mathbb Q_2$.

\tbl{Next, suppose $\mathbb Q_1$ and $\mathbb Q_2$ has the relation \eqref{Eq:AC}.} For an increasing function $u$, the expected utility is defined as $\mathbb E_u^{\mathbb Q}: X\mapsto \mathbb E^{\mathbb Q}(u(X))$. 
Instead of finding a simplified version of Theorem \ref{Thone} for this case, we apply Theorem \ref{Thone} directly to some specific $\rho$. For $\alpha\in (0,1)$, let
$\mathrm{LES}_{\alpha}^{\mathbb Q}(X)=\frac{1}{\alpha}\int_{1-\alpha}^{1}\VaR_t^{\mathbb Q}(X)\d t$, which represents the lower conditional tail expectation.

\begin{corollary}\label{Cor3} For $\Lambda\in\mathcal H$ with $0<\lambda^-\leq \lambda^+<1$,  we have the following conclusions.
\begin{enumerate}[(i)]
\item For $\alpha\in (0,1)$,  $$\Lambda\VaR^{\mathbb Q_1}\dsquare \ES_{\alpha}^{\mathbb Q_2}(X)=\inf_{t\in\R}\left(t+\frac{1-\lambda^+}{\alpha}\mathrm{LES}_{1-\lambda^+}^{\mathbb Q_1}(\eta(X-t)_+)\right);$$
\item For \tbl{$\X=L^\infty$} and an increasing function $u$, we have  \begin{align*}
&\Lambda\VaR^{\mathbb Q_1}\dsquare \mathbb E_u^{\mathbb Q_2}(X)\\
&=\inf_{x\in\R}\left(x+u(-\infty)(\mathbb Q_2(N)+\Lambda(x)\ES_{\Lambda(x)}^{\mathbb Q_1}(\eta))+\mathbb E^{\mathbb Q_1}(\eta u(X-x)\id_{\{U_\eta^{\mathbb Q_1}<1-\Lambda(x)\}})\right),
\end{align*}
    where $u(-\infty)=\lim_{y\to-\infty} u(y)$.
\end{enumerate}
\end{corollary}

Note that in (i) of Corollary  \ref{Cor3}, if $\Lambda$ is a constant, then the result coincides with Theorem 5 of \cite{ELMW19} for $n=2$, where the semi-explicit formula is offered. Using \eqref{Eq:AC}, it might be difficult to find more explicit expression for the inf-convolution of $\Lambda\VaR^{\mathbb Q_1}$ and $\rho_g^{\mathbb Q_2}$/$\Lambda_1\VaR^{+,\mathbb Q_2}$.


\section{Inf-convolution of $\Lambda\VaR^+$ and a monotone risk measure}\label{sec:4}
The inf-convolution of multiple $\Lambda\VaR^+$ under heterogeneous beliefs is a very challenging problem even if we suppose the beliefs are homogeneous; see the discussion for homogeneous beliefs in \cite{L24}. One important feature of $\Lambda\VaR^+$ is  to capture the tail risk for decreasing $\Lambda$; see \cite{FMP14} and \cite{HMP18}. In this section, we focus on the inf-convolution of $\Lambda\VaR^{+}$ and a monotone risk measure $\rho$ under $\mathbb Q_1$ and $\mathbb Q_2$, which includes the inf-convolution of two $\Lambda\VaR^+$ as a special case. 
For any $\Lambda\in\mathcal H$, we denote $\Lambda_{x}(z)=\inf_{x\leq t\leq z}\Lambda(t),~z\geq x$. Then $\Lambda_{x}(z)$ is decreasing and right-continuous for $z\geq x$.  If $\Lambda$ is decreasing and right-continuous, then $\Lambda_{x}(z)=\Lambda(z),~~z\geq x$; and if $\Lambda$ is increasing, then $\Lambda_{x}(z)=\Lambda(x),~~z\geq x$. The following result is valid for general $\Lambda\in\mathcal H$. 
We say $\rho$ is \emph{continuous from above} on $\X$ if  $X_n\downarrow X$ implies $\lim_{n\to\infty}\rho(X_n)=\rho(X)$ for $X_n, X\in \X$.
\begin{theorem}\label{Thfour}
Suppose $\Lambda\in\mathcal H$ with $0<\lambda^-\leq \lambda^+<1$,  and  $\rho$ is  monotone and law-invariant under $\mathbb Q_2$. \tbl{If either (i) $\X=L^\infty$ or (ii) $\X$ is unbounded and $\rho$ is continuous from above}, 
then we have
\begin{align}\label{Minimizer2}\Lambda\VaR^{+,\mathbb Q_1}\dsquare \rho^{\mathbb Q_2}(X)=\inf_{x\in\R}\inf_{y\geq x}\inf_{U\overset{\mathbb Q_1}{\sim} U[0,1]}\left\{x+\rho^{\mathbb Q_2}\left(X-\Lambda_{x,y}^{-1}(U)\right)\right\},
\end{align}
where
$$\Lambda_{x,y}(z)=\left\{\begin{array}{cc}
0,& z<x\\
1-\Lambda_x(z),& x\leq z<y\\
1,&z\geq y
\end{array}\right..$$
Moreover, for case (i), the optimal allocation of the inf-convolution exists if and only if the minimizer of \eqref{Minimizer2} exists. For both cases (i) and (ii),  if $(x^*,y^*, U^*)$ is the minimizer, then one optimal allocation is given by
\begin{align}\label{SolutionD}
X_1^*=\Lambda_{x^*,y^*}^{-1}(U^*),~ X_2^*=X-\Lambda_{x^*,y^*}^{-1}(U^*).
\end{align}
\end{theorem}

In the case of $\rho=\Lambda_1\VaR^{+}$  for $\Lambda_1\in\mathcal H$ with $0<\lambda_1^-\leq \lambda_1^+<1$, it follows from Theorem \ref{Thfour} that the optimal risk allocation for risk sharing between $\Lambda\VaR^{+}$ and  $\Lambda_1\VaR^{+}$ under $\mathbb Q_1$ and $\mathbb Q_2$ has the form of \eqref{SolutionD}, which is neither comonotonic nor pairwise countermonotonic. This means that the optimal risk sharing for multiple $\Lambda\VaR^{+}$ under heterogeneous beliefs is qualitatively different  from the cases with convex risk measures or quantile-based risk measures in the literature.
The expression of the inf-convolution in \eqref{Minimizer2} is an optimization problem involving three parameters $(x,y,U)$, where $U\sim U[0,1]$ under $\mathbb Q_1$. Note that the distribution of $U$ under $\mathbb Q_2$ depends on the relation between $\mathbb Q_1$ and $\mathbb Q_2$.   Moreover, the expression in \eqref{Minimizer2} and also the optimal allocation in \eqref{SolutionD} show that shape of $\Lambda$ function plays an important role in both the expression of the inf-convolution and the form of the optimal allocations, which is very different from the results of $\Lambda\VaR$ displayed in Sections \ref{sec:multiple} and \ref{sec:5}.

\tbl{For case (ii) of Theorem \ref{Thfour}, we suppose that $\rho$ is continuous from above. In fact, many risk functionals satisfy this assumption, such as  distortion risk measures $\rho_g$ with continuous $g$ (including $\ES_\alpha$ as a special case), expected utility $\mathbb E_u$ for continuous and increasing $u$, and the right quantiles. Later, we will define an alternative Lambda-$\VaR$ measure that also satisfies this assumption.}

 We are also aware that the conclusion in Theorem \ref{Thfour} is complicated in the sense that it involves $\inf_{U\overset{\mathbb Q_1}{\sim}U[0,1]}\rho^{\mathbb Q_2}(X-\Lambda_{x,y}^{-1}(U))$, which is a very difficult problem even if $\mathbb Q_1=\mathbb Q_2$. For $\mathbb Q_1=\mathbb Q_2$, $\inf_{U\overset{\mathbb Q_1}{\sim}U[0,1]}\rho^{\mathbb Q_2}(X-\Lambda_{x,y}^{-1}(U))$ corresponds to the problem of robust risk aggregation with dependence uncertainty, i.e., finding the value of $\inf_{X\sim F, Y\sim G}\rho(X+Y)$, where the marginal distributions are known but the dependence structure is completely unknown. We refer to \cite{WW16}, \cite{JHW16}, \cite{BLLW20}, \cite{HL24} and the references therein for the discussion on the robust risk aggregation.
 
\tbl{
We denote $X\leq_{\mathrm icx}Y$ if $\mathbb E(f (X))\leq\mathbb E(f (Y))$ for all increasing and convex function $f$. A mapping $\rho: \X \to \R$ is said to be SSD-consistent if $X\leq_{\mathrm icx}Y$ implies $\rho(X)\leq \rho(Y)$. Note that here $\X\subset L^1$.
\begin{corollary} Suppose $\X\subset L^1$.
 If $\mathbb Q_1=\mathbb Q_2=\mathbb Q$ and $\rho$ is SSD-consistent risk measure, then under the assumption of Theorem \ref{Thfour}, we have
\begin{align*}\Lambda\VaR^{+,\mathbb Q}\dsquare \rho^{\mathbb Q}(X)=\inf_{x\in\R}\inf_{y\geq x}\left\{x+\rho^{\mathbb Q}\left(X-\Lambda_{x,y}^{-1}(U_X^{\mathbb Q})\right)\right\}.
\end{align*}
\end{corollary}
Note that the above corollary coincides with Theorem 5 of \cite{L24}. We next see an example that $\rho$ is not SSD-consistent.
}
\tbl{
For $\Lambda:\R\to [0,1]$ and an increasing $f:\R\to [0,1]$, define
$$q_{\Lambda}^+(f)=\inf\{x\in\R: f(x)>1-\Lambda(x)\},$$
where $\inf\emptyset =\infty$. Note that if $\Lambda$ is a constant and $f$ is a distribution function, then $q_{\Lambda}^+(f)$ is the right quantile. We refer to \cite{BP22} and \cite{HL24} for the properties and applications of $q_{\Lambda}^+$.
}
\tbl{
Define an alternative version of $\Lambda\VaR$ by $\overline{\Lambda\VaR}(X)=q_{\Lambda}^+(F_X)$ for $X\in\X$. One can easily show that $\overline{\Lambda\VaR}$ is continuous from above if $\Lambda\in \mathcal H$. Then applying Theorem \ref{Thfour}, we have the following result.
\begin{corollary}\label{Cor:-1} Suppose $\mathbb Q_1=\mathbb Q_2=\mathbb Q$, $\Lambda\in \mathcal H$ with $0<\lambda^-\leq \lambda^+<1$ is strictly decreasing and $\Lambda_1\in\mathcal H$ with $0<\lambda_1^-\leq \lambda_1^+<1$. Then  we have, for $X\in \X$ with continuous $F_X^{\mathbb Q,-1}$ over $(0,1)$, 
    \begin{align*}\Lambda\VaR^{+,\mathbb Q}\dsquare \overline{\Lambda_1\VaR}^{\mathbb Q} (X)=\inf_{x\in\R}\inf_{y\geq x}\left\{x+q_{\Lambda_1}^+(l_{x,y}^{-1})\right\},
\end{align*}
where 
$l_{x,y}(t)=\sup_{s\in (0, t]}\{F_X^{\mathbb Q,-1}(s)-\Lambda_{x,y}^{-1}(1-t+s)\},~t\in (0,1)$ and
$l_{x,y}^{-1}(z)=\inf\{w\in (0,1): l_{x,y}(w)\geq z\},~z\in \R$ with $\inf\emptyset=1$.
\end{corollary}
Corollary \ref{Cor:-1} requires that $F_X^{\mathbb Q,-1}$ is continuous over $(0,1)$, which is a technical assumption for the computation of $\inf_{U\overset{\mathbb Q}{\sim} U[0,1]}\overline{\Lambda_1\VaR}^{\mathbb Q}\left(X-\Lambda_{x,y}^{-1}(U)\right)$. Our computation may not be correct if this assumption is violated.}

 In order to simplify \eqref{Minimizer2}, we consider a special case of $\Lambda$ functions. This makes it easier to find more explicit expressions of the inf-convolution for concrete examples of $\rho$.

\begin{proposition}\label{prop:step}
For $x_1\in\R$ and $0<\lambda_1<\lambda_2<1$, let $\Lambda(z)=(1-\lambda_1)\id_{\{z<x_1\}}+(1-\lambda_2)\id_{\{z\geq x_1\}}$. Under the assumption of Theorem \ref{Thfour}, we have
\begin{align}\label{Minimizer3}\Lambda\VaR^{+,\mathbb Q_1}\dsquare \rho^{\mathbb Q_2}(X)&=\inf_{x\in\R}\inf_{y\geq x\vee x_1}\inf_{(B_1,B_2)\in\mathcal B_{\lambda_1,\lambda_2}}\left\{x+\rho^{\mathbb Q_2}\left(X-x\id_{B_1}-x\vee x_1\id_{B_2}-y\id_{(B_1\cup B_2)^c}\right)\right\},
\end{align}
where $\mathcal B_{\lambda_1,\lambda_2}=\{(B_1,B_2): \mathbb Q_1(B_1)=\lambda_1, \mathbb  Q_1(B_2)=\lambda_2-\lambda_1, B_1\cap B_2=\emptyset\}$.
\end{proposition}
Note that the optimization problem in \eqref{Minimizer3} is much easier than that in \eqref{Minimizer2} because it only has parameters $(x,y,B_1,B_2)$. \tbl{To illustrate our Proposition \ref{prop:step}, we next see the example of $\rho^{\mathbb Q_2}=\ES_{\alpha}^{\mathbb Q_2}$. In the following corollary, we suppose $\mathbb Q_i(\cdot)=\mathbb P(\cdot|D_i)$ for some $D_i\in\mathcal F,~i=1,2$,  and $\mathbb Q(\cdot)=\mathbb P(\cdot|D_1\cap D_2)$  with $\mathbb P(D_1\cap D_2)>0$.
\begin{corollary}\label{Cor:-2}
For $x_1\in\R$ and $0<\lambda_1<\lambda_2<1$, let $\Lambda(z)=(1-\lambda_1)\id_{\{z<x_1\}}+(1-\lambda_2)\id_{\{z\geq x_1\}}$. Then we have
$$\Lambda\VaR^{+,\mathbb Q_1}\dsquare \ES_{\alpha}^{\mathbb Q_2}(X)=\left\{\begin{array}{cc}
\frac{1}{\alpha}\frac{\mathbb P(D_1\cap D_2)}{\mathbb P(D_2)}\int_{\alpha_0}^{\alpha_1}\VaR_{t}^{\mathbb Q}(X)\d t,&\lambda_1\mathbb P(D_1)-\mathbb P(D_1\cap D_2^c)\geq \alpha\mathbb P(D_2)\\
-\infty,& \text{otherwise}
\end{array}\right.,$$
where $\alpha_0=1-\frac{\lambda_1\mathbb P(D_1)-\mathbb P(D_1\cap D_2^c)}{\mathbb P(D_1\cap D_2)}$ and $\alpha_1=1-\frac{\lambda_1\mathbb P(D_1)-\mathbb P(D_1\cap D_2^c)-\alpha\mathbb P(D_2)}{\mathbb P(D_1\cap D_2)}$.
\end{corollary} }
\section{Conclusion}
In this paper, we consider the inf-convolution of multiple $\Lambda\VaR$  under heterogeneous beliefs. We obtain the general expression for the inf-convolution and the corresponding optimal allocations. To study impact of the relation of the beliefs on the inf-convolution,  we discuss three different cases: i) homogeneous beliefs; ii) conditional beliefs; iii) general beliefs with two agents. For all these cases, we obtain more explicit expressions of the inf-convolution. In all above cases we demonstrate that trivial outcomes arise when both belief inconsistency and risk tolerance are high.
The inf-convolution of one $\Lambda\VaR/\Lambda\VaR^+$ and a general risk measure with belief heterogeneity  are also discussed.
There are still some unsolved problems, such as what the inf-convolution of multiple $\Lambda\VaR$ under general heterogeneous beliefs is, and what the inf-convolution of multiple $\Lambda\VaR^+$ under heterogeneous beliefs is, which deserves further study.
\subsection*{Acknowledgments}
The authors thank the editor, an associate editor, and two anonymous referees for their helpful comments.

\section*{Competing Interests}
The authors declare no competing interests.

\appendix
\section{Proof of all results in Sections \ref{sec:2}-\ref{sec:4}}\label{Appendix}
{\bf Proof of Proposition \ref{prop:general1}}. Note that there exists $A\in\mathcal F$ such that $\mathbb Q_1(A)=0$ and $\mathbb Q_2(A)=1$. It follows from the law-invariance of $\rho_1$ and $\rho_2$ and the constant preservation of $\rho_2$ that \begin{align*}
\rho_1^{\mathbb Q_1}\dsquare \rho_2^{\mathbb Q_2}(X)&\leq \rho_1^{\mathbb Q_1}(X-c\id_A)+\rho_2^{\mathbb Q_2}(c\id_A)\\
&=\rho_1^{\mathbb Q_1}(X)+\rho_2^{\mathbb Q_2}(c)= \rho_1^{\mathbb Q_1}(X)+c\to-\infty
\end{align*}
as $c\to-\infty$. Hence, $\rho_1^{\mathbb Q_1}\dsquare \rho_2^{\mathbb Q_2}(X)=-\infty.$  This completes the proof. \qed

{\bf Proof of Theorem \ref{Thmain}}.  We first show that $\dsquare_{i=1}^n\Lambda_i\VaR^{\mathbb Q_i}(X)\geq \Gamma_{\Lambda_1,\dots,\Lambda_n}(X)$. For $(X_1,\dots, X_n)\in \mathbb A_n(X)$, let $x=\sum_{i=1}^{n}\Lambda_i\VaR^{\mathbb Q_i}(X_i)$ and $y_i=\Lambda_i\VaR^{\mathbb Q_i}(X_i)$. Note that $0<\lambda_i^-\leq \lambda_i^+<1$ implies $y_i\in\R$ and $x\in\R$. By the definition of $\Lambda_i\VaR$ and the right continuity of $\Lambda_i$, we have $\mathbb Q_i(C_i)\leq \Lambda_i(y_i)$ with $C_i=\{X_i>y_i\}$. We denote $D_i=C_i\cup \left(\cup_{i=1}^n C_i\right)^c$. Then it follows that
$$\{X>x\}=\left\{\sum_{i=1}^{n}X_i>\sum_{i=1}^{n}y_i\right\}\subset \bigcup_{i=1}^n C_i.$$ For $(A_1,\dots, A_n)\in\Pi_n(\Omega)$, let $A_i\subset D_i$. Then we have
$$\{X>x\}\cap A_i\subset\{X>x\}\cap D_i=\{X>x\}\cap C_i\subset C_i.$$
Using the fact $\mathbb Q_i(C_i)\leq \Lambda_i(y_i)$, we have $\mathbb Q_i(\{X>x\}\cap A_i)\leq \Lambda(y_i)$. This implies $\sum_{i=1}^{n}\Lambda_i\VaR^{\mathbb Q_i}(X_i)=x\geq \Gamma_{\Lambda_1,\dots,\Lambda_n}(X)$. By the arbitrary of $(X_1,\dots, X_n)\in \mathbb A_n(X)$, we have $\dsquare_{i=1}^n\Lambda_i\VaR^{\mathbb Q_i}(X)\geq \Gamma_{\Lambda_1,\dots,\Lambda_n}(X)$.

Next we show the inverse inequality. Suppose there exist $y_i\in\R, (A_1,\dots, A_n)\in\Pi_n(\Omega)$ such that $\mathbb Q_i\left(\{X>\sum_{j=1}^{n}y_j\}\cap A_i\right)\leq \Lambda_i(y_i)$. Let $X_i=(X-x)\id_{A_i}+y_i,~i=1,\dots, n$, where $x=\sum_{i=1}^{n}y_i$. Then $(X_1,\dots, X_n)\in \mathbb A_n(X)$. Direct computation gives $$\Lambda_i\VaR^{\mathbb Q_i}(X_i)=y_i+\Lambda_i^{y_i}\VaR^{\mathbb Q_i}((X-x)\id_{A_i}),$$ where $\Lambda_i^{y_i}(z)=\Lambda_i(z+y_i)$ for $z\in\R$. Note that $\mathbb Q_i((X-x)\id_{A_i}>0)=\mathbb Q_i(\{X>x\}\cap A_i)\leq \Lambda_i(y_i)=\Lambda_i^{y_i}(0)$. By definition, we have $\Lambda_i^{y_i}\VaR^{\mathbb Q_i}((X-x)\id_{A_i})\leq 0$. Hence,
$$\sum_{i=1}^{n}\Lambda_i\VaR^{\mathbb Q_i}(X_i)\leq \sum_{i=1}^{n} y_i=x.$$
This implies $\dsquare_{i=1}^n\Lambda_i\VaR^{\mathbb Q_i}(X)\leq \sum_{i=1}^{n}\Lambda_i\VaR^{\mathbb Q_i}(X_i)\leq \sum_{i=1}^{n}y_i$. By the arbitrary of $y_i$, we have $\dsquare_{i=1}^n\Lambda_i\VaR^{\mathbb Q_i}(X)\leq \Gamma_{\Lambda_1,\dots,\Lambda_n}(X)$. One can directly check that $X_i^*,~i=1,\dots,n$ is the optimal allocation. We complete the proof. \qed

{\bf Proof of Proposition \ref{prop:finite1}}. Case (i). Note that $\bigvee_{i=1}^n\frac{\mathbb Q_i(A_i)}{\lambda_i^-}\leq 1$ for some $(A_1,\dots, A_n)\in\Pi_n(\Omega)$ implies  $\mathbb Q_i(A_i)\leq \lambda_i^-$ for all $i=1,\dots,n$. Hence, for any $(y_1,\dots,y_n)$, it follows that $\mathbb Q_i\left(\{X>\sum_{j=1}^{n}y_j\}\cap A_i\right)\leq \mathbb Q_i(A_i)\leq \lambda_i^-\leq \Lambda_i(y_i)$. Consequently,  $\Gamma_{\Lambda_1,\dots,\Lambda_n}(X)=-\infty$.

Case (ii). \tbl{ We assume by contradiction that  $\Gamma_{\Lambda_1,\dots,\Lambda_n}(X)=-\infty$. By definition,  there exist $(y_1^{(m)},\dots,y_n^{(m)})$ with $\lim_{m\to\infty}\sum_{i=1}^{n}y_i^{(m)}=-\infty$ and $(A_1^{(m)},\dots, A_n^{(m)})\in\Pi_n(\Omega)$ such that  $\mathbb Q_i(\{X>\sum_{j=1}^{n}y_j^{(m)}\}\cap A_i^{(m)})\leq \Lambda_i(y_i^{(m)})$ for $i=1,\dots,n$. Moreover, using the condition $\inf_{(A_1,\dots, A_n)\in\Pi_n(\Omega)}\bigvee_{i=1}^n\frac{\mathbb Q_i(A_i)}{\lambda_i^+}>1$,  there exists $\epsilon>0$ such that $\bigvee_{i=1}^n\frac{\mathbb Q_i(A_i)}{\lambda_i^+}>1+\epsilon$ for any $(A_1,\dots, A_n)\in \Pi_n(\Omega)$. Hence, there exists $i_m\in \{1,2,\dots,n\}$ such that $\mathbb Q_{i_m}(A_{i_m}^{(m)})>(1+\epsilon)\lambda_{i_m}^+$.  For $\{i_m,m\geq 1\}$, we can find a constant subsequence. Without loss of generality, we assume $i_m=1$ for all $m\geq 1$. This means $\mathbb Q_{1}(A_1^{(m)})>(1+\epsilon)\lambda_1^+$. This together with the fact that $\lim_{m\to\infty}\sum_{i=1}^{n}y_i^{(m)}=-\infty$  implies for $m$ large enough, \begin{align*}\mathbb Q_1\left(\left\{X>\sum_{j=1}^{n}y_j^{(m)}\right\}\cap A_1^{(m)}\right)&=\mathbb Q_1(A_1^{(m)})-\mathbb Q_1\left(\left\{X\leq \sum_{j=1}^{n}y_j^{(m)}\right\}\cap A_1^{(m)}\right)\\
&>(1+\epsilon)\lambda_1^+-\mathbb Q_1\left(X\leq\sum_{j=1}^{n}y_j^{(m)}\right)\\
&\geq (1+\epsilon/2)\lambda_1^+,
\end{align*}
which contradicts $\mathbb Q_1(\{X>\sum_{j=1}^{n}y_j^{(m)}\}\cap A_1^{(m)})\leq \Lambda_1(y_1^{(m)})\leq \lambda_1^+$.
Hence,  $\Gamma_{\Lambda_1,\dots,\Lambda_n}(X)>-\infty$.} \qed

{\bf Proof of Proposition \ref{prop:finite2}}. We first consider the case that $\Lambda_i\in\mathcal H$ is decreasing. If $\bigvee_{i=1}^n\frac{\mathbb Q_i(A_i)}{\lambda_i^+}<1$ for some $(A_1,\dots, A_n)\in\Pi_n(\Omega)$, then $\mathbb Q_i(A_i)<\lambda_i^+$ for all $i=1,\dots,n$. Note that $\lim_{y_i\to-\infty}\Lambda(y_i)=\lambda_i^+$. Hence, there exists $y_i^0$ such that $\Lambda(y_i)\geq \mathbb Q_i(A_i)$ for all $y_i<y_i^0$. It follows that $\mathbb Q_i(\{X>\sum_{j=1}^{n} y_j\}\cap A_i)\leq \mathbb Q_i(A_i)\leq \Lambda_i(y_i)$ if $y_i\leq y_i^0$ for all $i=1,\dots,n$. This implies  $\Gamma_{\Lambda_1,\dots,\Lambda_n}(X)=-\infty$. The second statement is shown in (ii) of Proposition \ref{prop:finite1}.

Next, we consider the case  that $\Lambda_i\in\mathcal H$ is increasing. If $\bigwedge_{i=1}^n\left(\frac{\mathbb Q_i(A_i)}{\lambda_i^-}\vee\bigvee_{j\neq i}\frac{\mathbb Q_j(A_j)}{\lambda_j^+}\right)<1$ for some $(A_1,\dots, A_n)\in\Pi_n(\Omega)$, then there exists $i\in\{1,\dots,n\}$ such that
$\mathbb Q_i(A_i)<\lambda_i^-$ and $\mathbb Q_j(A_j)<\lambda_j^+$ for $j\neq i$. Note that there exists $y_j^0\in\R$ such that $\mathbb Q_j(A_j)\leq \Lambda_j(y_{j}^0)$ for all $j\neq i$. Moreover, for any $y_i\in\R$, it follows that $\mathbb Q_j(\{X>y_i+\sum_{k\neq i} y_k^0\}\cap A_j)\leq \mathbb Q_j(A_j)\leq \Lambda_j(y_j^0)$ for $j\neq i$, and $\mathbb Q_i(\{X>y_i+\sum_{k\neq i} y_k^0\}\cap A_i)\leq \Lambda_i(y_i)$. By the arbitrary of $y_i$, we have $\Gamma_{\Lambda_1,\dots,\Lambda_n}(X)=-\infty$.

We next show the last statement. 
Let $1+\epsilon=\inf_{(A_1,\dots, A_n)\in\Pi_n(\Omega)}\bigwedge_{i=1}^n\left(\frac{\mathbb Q_i(A_i)}{\lambda_i^-}\vee\bigvee_{j\neq i}\frac{\mathbb Q_j(A_j)}{\lambda_j^+}\right)$ for some $\epsilon>0$.
Suppose by contradiction that $\Gamma_{\Lambda_1,\dots,\Lambda_n}(X)=-\infty$. Then there exist $(y_1^{(m)},\cdots, y_n^{(m)})$ satisfying $\lim_{m\to\infty}\sum_{i=1}^{n}y_i^{(m)}=-\infty$, and  $(A_1^{(m)},\dots, A_n^{(m)})\in\Pi_n(\Omega)$ such that $\mathbb Q_i(\{X>\sum_{j=1}^n y_j^{(m)}\}\cap A_i^{(m)})\leq \Lambda_i(y_i^{(m)})$. Without loss of generality, we can assume  $\lim_{m\to\infty}y_i^{(m)}=y_i$. Let $E_1=\{i\in\{1,\dots, n\}: y_i=-\infty\}$ and $E_2=\{i\in\{1,\dots, n\}: y_i>-\infty\}$.   By the fact that $\lim_{m\to\infty}\sum_{i=1}^{n}y_i^{(m)}=-\infty$, we have $E_1\neq\emptyset$.  Moreover, for any $0<\eta<\epsilon$, 
there exists $m_0>0$ such that $\mathbb Q_i(X\leq\sum_{j=1}^n y_j^{(m)})<\lambda_i^-\eta/2$ for all $m\geq m_0$. Hence, we have $\mathbb Q_i(A_i^{(m)})\leq (1+\eta/2)\Lambda_i(y_i^{(m)})$ for all $m\geq m_0$. Note that $\lim_{m\to\infty}\Lambda_i(y_i^{(m)})=\lambda_i^-$ for all $i\in E_1$. Hence, for any $0<\eta<\epsilon$, there exists $m_1\geq m_0$ such that $\mathbb Q_i(A_i^{(m)})\leq \lambda_i^-(1+\eta)$ for all $m\geq m_1$ and $i\in E_1$. Moreover, $\mathbb Q_i(A_i^{(m)})\leq \lambda_i^+(1+\eta)$ for all $m\geq m_1$ and $i\in E_2$. Combination of the those conclusion leads to $\bigwedge_{i=1}^n\left(\frac{\mathbb Q_i(A_i^{(m)})}{\lambda_i^-}\vee\bigvee_{j\neq i}\frac{\mathbb Q_j(A_j^{(m)})}{\lambda_j^+}\right)\leq 1+\eta<1+\epsilon$ for $m>m_1$, which is a contradiction of the assumption. Hence, $\Gamma_{\Lambda_1,\dots,\Lambda_n}(X)>-\infty$. We complete the proof. \qed

{\bf Proof of Theorem \ref{prop:homogeneous}}. By Theorem \ref{Thmain}, we have $$\dsquare_{i=1}^n\Lambda_i\VaR^{\mathbb P}(X)=\inf\left\{\sum_{j=1}^{n}y_j: \mathbb P\left(\left\{X>\sum_{j=1}^{n}y_j\right\}\cap A_i\right)\leq \Lambda_i(y_i)~\text{for some}~(A_1,\dots, A_n)\in\Pi_n(\Omega)\right\}.$$
Note that $\mathbb P\left(\{X>\sum_{j=1}^{n}y_j\}\cap A_i\right)\leq \Lambda_i(y_i)$ for some $(A_1,\dots, A_n)\in\Pi_n(\Omega)$ if and only if
$\mathbb P\left(X>\sum_{i=1}^{n}y_i\right)\leq \sum_{i=1}^{n}\Lambda_i(y_i)$. Hence, we have
\begin{align}\dsquare_{i=1}^n\Lambda_i\VaR^{\mathbb P}(X)&=\inf\left\{\sum_{i=1}^{n}y_i: (y_1,\dots,y_n)\in\R^n, \mathbb P\left(X>\sum_{i=1}^{n}y_i\right)\leq \sum_{i=1}^{n}\Lambda_i(y_i)\right\}\nonumber\\
&=\inf\left\{x\in\R: \sum_{i=1}^{n}y_i=x, \mathbb P\left(X>x\right)\leq \sum_{i=1}^{n}\Lambda_i(y_i)\right\}\label{eq:general}\\
&=\inf\left\{x\in\R: \mathbb P\left(X>x\right)\leq \Lambda^*(x)\right\}=\Lambda^*\VaR^{\mathbb P}(X).\nonumber
\end{align}
For the optimal allocation, note that the right continuity of $\Lambda^*$ at $x^*$ implies $\mathbb P\left(X>x^*\right)\leq \Lambda^*(x^*)$. Moreover, there exist  $(y_1^*, \dots, y_n^*)$ and $(A_1^*, \dots, A_n^*)$ such that $1\bigwedge\sum_{i=1}^{n}\Lambda_i(y_i^*)=\Lambda^*(x^*)$  and  $\mathbb P\left(\{X>x^*\}\cap A_i^*\right)\leq \Lambda_i(y_i^*)$. Hence, $(y_1^*, \dots, y_n^*, A_1^*, \dots, A_n^*)$ is the minimizer of $\Gamma_{\Lambda_1,\dots,\Lambda_n}(X)$. Using Theorem \ref{Thmain}, the claimed allocation is an optimal allocation. \qed

{\bf Proof of Proposition \ref{prop:General}}. In light of \eqref{eq:general}, we have \begin{align*}\dsquare_{i=1}^n\Lambda_i\VaR^{\mathbb P}(X)
&=\inf\left\{x\in\R: \sum_{i=1}^{n}y_i=x, \mathbb P\left(X>x\right)\leq \sum_{i=1}^{n}\Lambda_i(y_i)\right\}\\
&=\inf_{\mathbf y_{n-1}\in\R^{n-1}}\Lambda^{\mathbf{y}_{n-1}}\VaR^{\mathbb P}(X).
\end{align*} \qed

{\bf Proof of Theorem \ref{prop:heterogeneous}}. By Theorem \ref{Thmain}, we have \begin{align*}&\dsquare_{i=1}^n\Lambda_i\VaR^{\mathbb Q_i}(X)\\
&=\inf\left\{\sum_{j=1}^{n}y_j: \mathbb P\left(\left\{X>\sum_{j=1}^{n}y_j\right\}\cap B_i\cap A_i\right)\leq \mathbb P(B_i)\Lambda_i(y_i)~\text{for some}~(A_1,\dots, A_n)\in\Pi_n(\Omega)\right\}.
\end{align*}
Let $\mathcal N_n=\{\{i_1,\dots, i_m\}\subset\{1,\dots,n\}:~m=1,2,\dots, n-1\}$. For $\{i_1,\dots, i_m\}\in \mathcal N_n$, we denote $C_{\{i_1,\dots,i_m\}}=\left(\bigcap_{i\in \{i_1,\dots,i_m\}} B_i\right)\setminus\left(\bigcup_{i\in \{1,\dots,n\}\setminus \{i_1,\dots,i_m\}} B_i\right) $. Then for $\{i_1,\dots,i_m\}, \{i_1',\dots,i_{m'}'\}\in \mathcal N_n$, we have $C_{\{i_1,\dots,i_m\}}\cap C_{\{i_1',\dots,i_{m'}'\}}=\emptyset$ if $\{i_1,\dots,i_m\}\neq \{i_1',\dots,i_{m'}'\}$. Moreover, it follows that $\bigcup_{\{i_1,\dots,i_m\}\in \mathcal N_n}C_{\{i_1,\dots,i_m\}}=\left(\bigcup_{i=1}^n B_i\right)\setminus\left(\bigcap_{i=1}^n B_i\right)$. Next, for each $(A_1,\dots, A_n)\in\Pi_n(\Omega)$, we do some operations on it. We fix $\{i_1,\dots,i_m\}\in \mathcal N_n$. For each $i\in \{i_1,\dots,i_m\}$, we do the following operations: $$A_i\to A_i\setminus C_{\{i_1,\dots,i_m\}}, ~A_j\to A_j\cup \left(A_i\cap C_{\{i_1,\dots,i_m\}}\right),$$ where $j=\min(\{1,\dots,n\}\setminus \{i_1,\dots,i_m\})$. The new sets after the operations are denoted by $(A_1',\dots, A_n')$. Clearly, $(A_1',\dots, A_n')\in \Pi_n(\Omega)$ and $B_i\cap A_i'\subset B_i\cap A_i$ for all $i=1,\dots,n$ and  $A_i'\cap C_{\{i_1,\dots,i_m\}}=\emptyset$ for all $i\in \{i_1,\dots,i_m\}$. Those operations will be done for all $\{i_1,\dots,i_m\}\in \mathcal N_n$. The final sets are denoted by $(A_1'',\dots, A_n'')$. Clearly, $(A_1'',\dots, A_n'')\in \Pi_n(\Omega)$ and $B_i\cap A_i''\subset B_i\cap A_i$. Moreover, it follows that $A_i''\cap C_{\{i_1,\dots,i_m\}}=\emptyset$ for all $\{i_1,\dots,i_m\}\in\mathcal N_n$ satisfying $i\in \{i_1,\dots,i_m\}$. This implies $B_i\cap A_i''=\left(\cap_{i=1}^n B_i\right)\cap A_i''$ for all $i=1,\dots,n$. By the construction of $A_i''$, it follows that $\left(\cap_{i=1}^n B_i\right)\cap A_i\subset B_i\cap A_i''$, which together with the fact $B_i\cap A_i''\subset B_i\cap A_i$ implies $\left(\cap_{i=1}^n B_i\right)\cap A_i=\left(\cap_{i=1}^n B_i\right)\cap A_i''=B_i\cap A_i''$.  Hence, we have
\begin{align*}
&\inf\left\{\sum_{j=1}^{n}y_j: \mathbb P\left(\left\{X>\sum_{j=1}^{n}y_j\right\}\cap B_i\cap A_i\right)\leq \mathbb P(B_i)\Lambda_i(y_i)~\text{for some}~(A_1,\dots, A_n)\in\Pi_n(\Omega)\right\}\\
&\leq \inf\left\{\sum_{j=1}^{n}y_j: \mathbb P\left(\left\{X>\sum_{j=1}^{n}y_j\right\} \cap\left(\cap_{j=1}^n B_j\right)\cap A_i\right)\leq \mathbb P(B_i)\Lambda_i(y_i)~\text{for some}~(A_1,\dots, A_n)\in\Pi_n(\Omega)\right\}.
\end{align*}
Note that the inverse inequality is trivial. Hence, we have
\begin{align*}
&\dsquare_{i=1}^n\Lambda_i\VaR^{\mathbb Q_i}(X)\\
&=\inf\left\{\sum_{j=1}^{n}y_j: \mathbb P\left(\left\{X>\sum_{j=1}^{n}y_j\right\}\cap \left(\cap_{j=1}^n B_j\right)\cap A_i\right)\leq \mathbb P(B_i)\Lambda_i(y_i)~\text{for some}~(A_1,\dots, A_n)\in\Pi_n(\Omega)\right\}.
\end{align*}
Clearly, $\dsquare_{i=1}^n\Lambda_i\VaR^{\mathbb Q_i}(X)=-\infty$ if $\mathbb P\left(\cap_{i=1}^n B_i\right)=0$. Next, we consider the case $\mathbb P\left(\cap_{i=1}^n B_i\right)>0$.
Note that $\mathbb P\left(\{X>\sum_{j=1}^{n}y_j\}\cap \left(\cap_{j=1}^n B_j\right)\cap A_i\right)\leq \mathbb P(B_i)\Lambda_i(y_i)$ for some $(A_1,\dots, A_n)\in\Pi_n(\Omega)$ and some $(y_1,\dots,y_n)\in\mathbb R^n$ is equivalent to  $\mathbb P\left(\{X>\sum_{j=1}^{n}y_j\}\cap (\cap_{j=1}^n B_j)\right)\leq \sum_{i=1}^{n}\mathbb P(B_i)\Lambda_i(y_i)$ for some $(y_1,\dots,y_n)\in\mathbb R^n$.
Using this conclusion and the attainability of $\Lambda^\diamond$,  we have
\begin{align*}
\dsquare_{i=1}^n\Lambda_i\VaR^{\mathbb Q_i}(X)
&=\inf\left\{x: x=\sum_{i=1}^{n}y_i, \mathbb P\left(\{X>x\}\cap(\cap_{j=1}^n B_j)\right)\leq \sum_{i=1}^{n}\mathbb P(B_i)\Lambda_i(y_i)\right\}\\
&=\inf\left\{x: x=\sum_{i=1}^{n}y_i, \mathbb Q\left(X>x\right)\leq \sum_{i=1}^{n}\frac{\mathbb P(B_i)\Lambda_i(y_i)}{\mathbb P(\cap_{j=1}^n B_j)}\right\}\\
&=\inf\left\{x: \mathbb Q\left(X>x\right)\leq \Lambda^\diamond(x)\right\}=\Lambda^\diamond\VaR^{\mathbb Q}(X).
\end{align*}
One can easily check that the claimed allocation is optimal. \qed

{\bf Proof of Theorem \ref{prop:continuous}}. Using Theorem \ref{Thmain}, we have  \tbl{\begin{align*}&\Lambda_1\VaR^{\mathbb Q_1}\dsquare \Lambda_2\VaR^{\mathbb Q_2}(X)\\
&=\inf\left\{x\in\R: \mathbb Q_1\left(\{X>x\}\cap A^c\right)\leq \Lambda_1(y), \mathbb Q_2\left(\{X>x\}\cap A\right)\leq \Lambda_2(x-y),~\text{for some}~y\in\R, A\in\mathcal F\right\}\\
&=\inf\left\{x\in\R: \mathbb Q_1\left(\{X>x\}\cap A^c\right)\leq \Lambda_1(y), \mathbb E^{\mathbb Q_1}(\eta\id_{\{X>x\}\cap A})+\mathbb Q_2(\{X>x\}\cap A\cap N)\leq \Lambda_2(x-y),\right.\\
&~~~~~~~~~~~~\left.\text{for some}~y\in\R, A\in\mathcal F\right\}.
\end{align*}
Let $A'=A\cap N^c$. Using the fact $\mathbb Q_1(N)=0$, we have $ \mathbb Q_1\left(\{X>x\}\cap (A')^c\right)= \mathbb Q_1\left(\{X>x\}\cap A^c\right)$, $\mathbb E^{\mathbb Q_1}(\eta\id_{\{X>x\}\cap A'})=\mathbb E^{\mathbb Q_1}(\eta\id_{\{X>x\}\cap A})$ and $\mathbb Q_2(\{X>x\}\cap A'\cap N)=0$. Hence, the inf-convolution can be simplified as 
\begin{align*}&\Lambda_1\VaR^{\mathbb Q_1}\dsquare \Lambda_2\VaR^{\mathbb Q_2}(X)\\
&=\inf\left\{x\in\R: \mathbb Q_1\left(\{X>x\}\cap A^c\right)\leq \Lambda_1(y), \mathbb E^{\mathbb Q_1}(\eta\id_{\{X>x\}\cap A})\leq \Lambda_2(x-y),~\text{for some}~y\in\R, A\in\mathcal F\right\}.
\end{align*}}
 Let $f(t)=\mathbb Q_1(\{X>x\}\cap\{U_\eta^{\mathbb Q_1}\leq t\}),~t\in [0,1]$. Clearly, $f$ is an increasing and continuous function. Let $f^{-1}$ denote the left-quantile of $f$. Define $A_{x,t}=\{U_\eta^{\mathbb Q_1}\leq f^{-1}(t)\}$ for $t\in [0,1-F_X^{\mathbb Q_1}(x)]$. Let $t_0=\mathbb Q_1(\{X>x\}\cap A)$. Then it follows that $\mathbb Q_1(\{X>x\}\cap A_{x,t_0})=\mathbb Q_1(\{X>x\}\cap A)$ and $$\mathbb E^{\mathbb Q_1}(\eta\id_{\{X>x\}\cap A_{x,t_0}})=\mathbb E^{\mathbb Q_1}(F_{\eta}^{\mathbb Q_1, -1}(U_\eta^{\mathbb Q_1})\id_{\{X>x\}\cap A_{x,t_0}})\leq \mathbb E^{\mathbb Q_1}(\eta\id_{\{X>x\}\cap A}).$$
Consequently, we have \begin{align*}&\Lambda_1\VaR^{\mathbb Q_1}\dsquare \Lambda_2\VaR^{\mathbb Q_2}(X)\\
&=\inf\left\{x\in\R: \mathbb Q_1\left(\{X>x\}\cap A_{x,t}^c\right)\leq \Lambda_1(y), \mathbb E^{\mathbb Q_1}(\eta\id_{\{X>x\}\cap A_{x,t}})\leq \Lambda_2(x-y),\right.\nonumber\\
&~~~~~~~~~~~~\left.\text{for some}~y\in\R, t\in [0,1-F_X^{\mathbb Q_1}(x)]\right\}\\
&=\inf\left\{x\in\R: 1-F_X^{\mathbb Q_1}(x)-t\leq \Lambda_1(y), g_{x}(t)\leq \Lambda_2(x-y),~\text{for some}~y\in\R, t\in [0,1-F_X^{\mathbb Q_1}(x)]\right\}\\
&=\inf\left\{x\in\R: g_{x}((1-F_X^{\mathbb Q_1}(x)-\Lambda_1(y))_+)\leq \Lambda_2(x-y),~\text{for some}~y\in\R\right\}.
\end{align*}
Next, we show $(X_1, X_2)$ given by \eqref{OptimalAb} is the optimal allocation. Note that 
\begin{align*}
\mathbb Q_1(X_1>y^*)&=\mathbb Q_1(\{X>x^*\}\cap (\Omega\setminus A^*))=\mathbb Q_1(X>x^*)-\mathbb Q_1(\{X>x^*\}\cap A^*)\\
&=1-F_X^{\mathbb Q_1}(x^*)-(1-F_X^{\mathbb Q_1}(x^*)-\Lambda_1(y^*))_+\leq \Lambda_1(y^*),\\
\mathbb Q_2(X_2>x^*-y^*)&=\mathbb Q_2(\{X>x^*\}\cap A^*)=\mathbb E^{\mathbb Q_1}\left(\eta\id_{\{X>x^*\}\cap \{U_\eta^{\mathbb Q_1}\leq z^*\}}\right)\\
&=g_{x^*}((1-F_X^{\mathbb Q_1}(x^*)-\Lambda_1(y^*))_+)\leq \Lambda_2(x^*-y^*).
\end{align*}
Hence, we have $\Lambda_1\VaR^{\mathbb Q_1}(X_1)\leq y^*$ and $\Lambda_2\VaR^{\mathbb Q_2}(X_2)\leq x^*-y^*$, which implies $\Lambda_1\VaR^{\mathbb Q_1}(X_1)+\Lambda_2\VaR^{\mathbb Q_2}(X_2)\leq \Lambda_1\VaR^{\mathbb Q_1}\dsquare \Lambda_2\VaR^{\mathbb Q_2}(X)$. By the definition of the inf-convolution,  $(X_1,X_2)$ is the optimal allocation.
This completes the proof. \qed

{\bf Proof of Corollary \ref{Cor:new}}. Note that $\mathbb Q_2(N)\geq 1-\lambda_2^-$ implies $ g_{x}(t)\leq 1-\mathbb Q_2(N)\leq \lambda_2^-\leq \Lambda_2(x-y)$ for all $x,y\in\R$. Hence, the inf-convolution is $-\infty$. \qed


{\bf Proof of Theorem \ref{Thone}}.
For any $x,y$ and $B$ satisfying $\mathbb Q_1(B)=1-\Lambda(x)$, let $X_1=x\id_{B}+(X-y)\id_{B^c}$.  Direct calculation gives
  $$\Lambda\VaR^{\mathbb Q_1}(X_1)+\rho^{\mathbb Q_2}(X-X_1)\leq x+\rho^{\mathbb Q_2}(X-X_1)=x+\rho^{\mathbb Q_2}((X-x)\id_{B}+y\id_{B^c}).$$
    This implies $\Lambda\VaR^{\mathbb Q_1}\dsquare \rho^{\mathbb Q_2}(X)\leq x+\rho^{\mathbb Q_2}((X-x)\id_{B}+y\id_{B^c})$.  
By the arbitrary of $x, y$ and $B$ satisfying $\mathbb Q_1(B)=1-\Lambda(x)$, we have $$\Lambda\VaR^{\mathbb Q_1}\dsquare \rho^{\mathbb Q_2}(X)\leq \inf_{x\in\R}\inf_{y\in\R}\inf_{\mathbb Q_1(B)=1-\Lambda(x)}\left\{x+\rho^{\mathbb Q_2}((X-x)\id_{B}+y\id_{B^c})\right\}.$$

 We next show the inverse inequality.
  For any $X_1\in\X$, let $x_1=\Lambda\VaR^{\mathbb Q_1}(X_1)$. Note that $X_1=F_{X_1}^{\mathbb Q_1, -1}(U_{X_1}^{\mathbb Q_1})$ a.s. under $\mathbb Q_1$.   Define $Y_1=x_1\id_{\{U_{X_1}^{\mathbb Q_1}\leq 1-\Lambda(x_1)\}}+(X+m)\id_{\{U_{X_1}^{\mathbb Q_1}>1-\Lambda(x_1)\}}$ with $m>m_0$.  
  Note that $U_{X_1}^{\mathbb Q_1}$ can be chosen such that $\{U_{X_1}^{\mathbb Q_1}\leq t\}\subset\{X_1\leq F_{X_1}^{\mathbb Q_1, -1}(t)\}$ for all $t\in (0,1)$. By the definition of $\Lambda\VaR$ and the right continuity of $\Lambda$, we have $F_{X_1}^{\mathbb Q_1}(x_1)\geq 1-\Lambda(x_1)$, which implies $x_1\geq F_{X_1}^{\mathbb Q_1, -1}(1-\Lambda(x_1))$. Hence, $X_1\leq x_1$ for all $w\in \{U_{X_1}^{\mathbb Q_1}\leq 1-\Lambda(x_1)\}$.  
  
  For the case $\mathcal X=L^\infty$, let $m_0=(x_1-\essinf^{\mathbb Q_1}X)\vee (\esssup^{\mathbb Q_1}|X|+\esssup^{\mathbb Q_2} |X_1|+|x_1|)$, where $\essinf^{\mathbb Q_i}$ and $\esssup^{\mathbb Q_i}$ represent the $\essinf$ and $\esssup$ under $\mathbb Q_i$, $i=1,2$.
 Then we have 
 $X_1\leq Y_1$ a.s. under $\mathbb Q_2$, which further implies $X-X_1\geq X-Y_1$ a.s. under $\mathbb Q_2$. 
 By the monotonicity of $\rho$, we have
 \begin{align*}\Lambda\VaR^{\mathbb Q_1}(X_1)+\rho^{\mathbb Q_2}(X-X_1)&\geq x_1+\rho^{\mathbb Q_2}(X-Y_1)\\
 &=x_1+\rho^{\mathbb Q_2}\left((X-x_1)\id_{\{U_{X_1}^{\mathbb Q_1}\leq 1-\Lambda(x_1)\}}-m\id_{\{U_{X_1}^{\mathbb Q_1}>1-\Lambda(x_1)\}}\right)\\
 &\geq \inf_{x\in\R}\inf_{y\in\R}\inf_{\mathbb Q_1(B)=1-\Lambda(x)}\left\{x+\rho^{\mathbb Q_2}((X-x)\id_{B}+y\id_{B^c})\right\}.
 \end{align*}
\tbl{For the case that $\rho^{\mathbb Q_2}$ is an $\epsilon$-tail risk measure, let $m_0=-\VaR_{\epsilon}^{\mathbb Q_2}(X-X_1)$. Then we have for $x\geq \VaR_{\epsilon}^{\mathbb Q_2}(X-X_1)$
 \begin{align*}
 &\mathbb Q_2(X-X_1\leq x)\\
 &=\mathbb Q_2(\{X-X_1\leq x\}\cap \{U_{X_1}^{\mathbb Q_1}\leq 1-\Lambda(x_1)\})+\mathbb Q_2(\{X-X_1\leq x\}\cap \{U_{X_1}^{\mathbb Q_1}>1-\Lambda(x_1)\})\\
 &\leq\mathbb Q_2(\{X-x_1\leq x\}\cap \{U_{X_1}^{\mathbb Q_1}\leq 1-\Lambda(x_1)\})+\mathbb Q_2(U_{X_1}^{\mathbb Q_1}>1-\Lambda(x_1))\\
 &=\mathbb Q_2(X-Y_1\leq x),
 \end{align*}
 which implies
 $$\frac{(F_{X-X_1}^{\mathbb Q_2}(x)-(1-\epsilon))_+}{\epsilon}\leq \frac{(F_{X-Y_1}^{\mathbb Q_2}(x)-(1-\epsilon))_+}{\epsilon},~x\in\R.$$
 By the monotonicity of $\rho$ and the fact $\rho$ is an $\epsilon$-tail risk measure, we have}
 \begin{align*}\Lambda\VaR^{\mathbb Q_1}(X_1)+\rho^{\mathbb Q_2}(X-X_1)&\geq x_1+\rho^{\mathbb Q_2}(X-Y_1)\\
 &=x_1+\rho^{\mathbb Q_2}\left((X-x_1)\id_{\{U_{X_1}^{\mathbb Q_1}\leq 1-\Lambda(x_1)\}}-m\id_{\{U_{X_1}^{\mathbb Q_1}>1-\Lambda(x_1)\}}\right)\\
 &\geq \inf_{x\in\R}\inf_{y\in\R}\inf_{\mathbb Q_1(B)=1-\Lambda(x)}\left\{x+\rho^{\mathbb Q_2}((X-x)\id_{B}+y\id_{B^c})\right\}.
 \end{align*}
 By the arbitrary of $X_1$, we obtain $$\Lambda\VaR^{\mathbb Q_1}\dsquare \rho^{\mathbb Q_2}(X)\geq\inf_{x\in\R}\inf_{y\in\R}\inf_{\mathbb Q_1(B)=1-\Lambda(x)}\left\{x+\rho^{\mathbb Q_2}((X-x)\id_{B}+y\id_{B^c})\right\}.$$
 Combing the above conclusions, we have \eqref{Minimizer1} holds.

We next show that the existence of the optimal allocation implies the existence of the minimizer  of \eqref{Minimizer1}. Suppose  there exists $X_1\in\X$ such that $\Lambda\VaR^{\mathbb Q_1}(X_1)+\rho^{\mathbb Q_2}(X-X_1)=\Lambda\VaR^{\mathbb Q_1}\dsquare \rho^{\mathbb Q_2}(X)$. Following the same argument as above to show the inverse inequality, let $x_1=\Lambda\VaR^{\mathbb Q_1}(X_1)$ and $Y_1=x_1\id_{\{U_{X_1}^{\mathbb Q_1}\leq 1-\Lambda(x_1)\}}+(X+m)\id_{\{U_{X_1}^{\mathbb Q_1}>1-\Lambda(x_1)\}}$ with some $m>m_0$ such that $X_1\leq Y_1$ a.s. under $\mathbb Q_2$ if $\X=L^\infty$, and $(F_{X-X_1}^{\mathbb Q_2}(x)-(1-\epsilon))_+\leq (F_{X-Y_1}^{\mathbb Q_2}(x)-(1-\epsilon))_+,~x\in\R$ if $\X$ is unbounded and $\rho^{\mathbb Q_2}$ is an $\epsilon$-tail risk measure. We have  \begin{align*}\Lambda\VaR^{\mathbb Q_1}(X_1)+\rho^{\mathbb Q_2}(X-X_1)&\geq \Lambda\VaR^{\mathbb Q_1}(Y_1) +\rho^{\mathbb Q_2}(X-Y_1)\\
 &=x_1+\rho^{\mathbb Q_2}((X-x_1)\id_{\{U_{X_1}^{\mathbb Q_1}\leq 1-\Lambda(x_1)\}}-m\id_{\{U_{X_1}^{\mathbb Q_1}>1-\Lambda(x_1)\}})\\
 &\geq \inf_{x\in\R}\inf_{y\in\R}\inf_{\mathbb Q_1(B)=1-\Lambda(x)}\left\{x+\rho^{\mathbb Q_2}((X-x)\id_{B}+y\id_{B^c})\right\}.
 \end{align*} Using $\Lambda\VaR^{\mathbb Q_1}(X_1)+\rho^{\mathbb Q_2}(X-X_1)=\Lambda\VaR^{\mathbb Q_1}\dsquare \rho^{\mathbb Q_2}(X)$ and \eqref{Minimizer1}, we have $$x_1+\rho^{\mathbb Q_2}((X-x_1)\id_{\{U_{X_1}^{\mathbb Q_1}\leq 1-\Lambda(x_1)\}}-m\id_{\{U_{X_1}^{\mathbb Q_1}>1-\Lambda(x_1)\}})
 =\inf_{x\in\R}\inf_{y\in\R}\inf_{\mathbb Q_1(B)=1-\Lambda(x)}\left\{x+\rho^{\mathbb Q_2}((X-x)\id_{B}+y\id_{B^c})\right\}.$$ This implies $(x_1,-m, \{U_{X_1}^{\mathbb Q_1}\leq 1-\Lambda(x_1)\})$ is the minimizer of \eqref{Minimizer1}.

Now suppose $(x^*,y^*, B^*)$ is the minimizer of \eqref{Minimizer1}. We next check $(X_1^*,X_2^*)$ given in \eqref{Minimizer20} is an optimal allocation. It follows that
 \begin{align*}\Lambda\VaR^{\mathbb Q_1}(X_1^*)+\rho^{\mathbb Q_2}(X_2^*)&\leq x^*+\rho^{\mathbb Q_2}((X-x^*)\id_{B^*}+y^*\id_{(B^*)^c})\\
 &=\inf_{x\in\R}\inf_{y\in\R}\inf_{\mathbb Q_1(B)=1-\Lambda(x)}\left\{x+\rho^{\mathbb Q_2}((X-x)\id_{B}+y\id_{B^c})\right\}\\
 &=\Lambda\VaR^{\mathbb Q_1}\dsquare \rho^{\mathbb Q_2}(X).
 \end{align*}
 Consequently, $\Lambda\VaR^{\mathbb Q_1}(X_1^*)+\rho^{\mathbb Q_2}(X_2^*)=\Lambda\VaR^{\mathbb Q_1}\dsquare \rho^{\mathbb Q_2}(X)$ and $(X_1^*, X_2^*)$ is an optimal allocation. This completes the proof. \qed

{\bf Proof of Proposition \ref{Pro:conditional}}. In light of Theorem \ref{Thone}, we have
$$\Lambda\VaR^{\mathbb Q_1}\dsquare \rho^{\mathbb Q_2}(X)=\inf_{x\in\R}\inf_{y\in\R}\inf_{\mathbb Q_1(B)=1-\Lambda(x)}\left\{x+\rho^{\mathbb Q_2}((X-x)\id_{B}+y\id_{B^c})\right\}.$$

 If $\mathbb P(B_1\cap B_2^c)\geq (1-\Lambda(x))\mathbb P(B_1)$, then there exists $D\in\mathcal F$ such that $D\subseteq B_1\cap B_2^c$ and $\mathbb Q_1(D)=1-\Lambda(x)$.  Using the law-invariance of $\rho$ under $\mathbb Q_2$, we have
\begin{align*}
\inf_{y\in\R}\inf_{\mathbb Q_1(B)=1-\Lambda(x)}\left\{x+\rho^{\mathbb Q_2}((X-x)\id_{B}+y\id_{B^c})\right\}&\leq \inf_{y\in\R}\left\{x+\rho^{\mathbb Q_2}((X-x)\id_{D}+y\id_{D^c})\right\}\\
&=\inf_{y\in\R}\left\{x+\rho^{\mathbb Q_2}(y)\right\}.
\end{align*}

If $\X=L^\infty$, for $y<\essinf^{\mathbb Q_2}(X-x)$, we have $(X-x)\id_{B}+y\id_{B^c}\geq y$ a.s. under $\mathbb Q_2$. It follows from the monotonicity and law-invariance of $\rho$ that $\rho^{\mathbb Q_2}((X-x)\id_{B}+y\id_{B^c})\geq \rho^{\mathbb Q_2}(y)$. Using the monotonicity of $\rho$ again, we have $$\inf_{y\in\R}\inf_{\mathbb Q_1(B)=1-\Lambda(x)}\left\{x+\rho^{\mathbb Q_2}((X-x)\id_{B}+y\id_{B^c})\right\}\geq\inf_{y\in\R}\left\{x+\rho^{\mathbb Q_2}(y)\right\}.$$

\tbl{If $\rho^{\mathbb Q_2}$ is an $\epsilon$-tail risk measure, for $y<\VaR_\epsilon^{\mathbb Q_2}(X-x)$, we have for $z\geq y$
\begin{align*}\mathbb Q_2((X-x)\id_{B}+y\id_{B^c}\leq z)=\mathbb Q_2(\{X-x\leq z\}\cap B)+\mathbb Q_2(B^c)\leq 1=\mathbb Q_2(y\leq z).
\end{align*}
Moreover, $\mathbb Q_2((X-x)\id_{B}+y\id_{B^c}<y)=\mathbb Q_2(\{X-x<y\}\cap B)<1-\epsilon$. Hence, $\VaR_\epsilon^{\mathbb Q_2}((X-x)\id_{B}+y\id_{B^c})\geq y$ and 
$$\frac{(\mathbb Q_2((X-x)\id_{B}+y\id_{B^c}\leq z)-(1-\epsilon))_+}{\epsilon}\leq \frac{(\id_{[y,\infty)]}(z)-(1-\epsilon))_+}{\epsilon},~z\in\R.$$
Using the monotonicity of $\rho$ and the fact that $\rho$ is an $\epsilon$-tail risk measure under $\mathbb Q_2$, we have $$\inf_{y\in\R}\inf_{\mathbb Q_1(B)=1-\Lambda(x)}\left\{x+\rho^{\mathbb Q_2}((X-x)\id_{B}+y\id_{B^c})\right\}\geq\inf_{y\in\R}\left\{x+\rho^{\mathbb Q_2}(y)\right\}.$$}
Consequently,
$$\inf_{y\in\R}\inf_{\mathbb Q_1(B)=1-\Lambda(x)}\left\{x+\rho^{\mathbb Q_2}((X-x)\id_{B}+y\id_{B^c})\right\}=\inf_{y\in\R}\left\{x+\rho^{\mathbb Q_2}(y)\right\}.$$
Next, we consider the case $\mathbb P(B_1\cap B_2^c)<(1-\Lambda(x))\mathbb P(B_1)$. Let $A_x=(B_1\cap B_2^c)\cup(B_1\cap B_2\cap\{U_X^{\mathbb P}<\alpha_x\})$  with $\alpha_x$ satisfying $\mathbb P(B_1\cap B_2\cap\{U_X^{\mathbb P}<\alpha_x\})=(1-\Lambda(x))\mathbb P(B_1)-\mathbb P(B_1\cap B_2^c)$. For any $B\in\mathcal F$ with $\mathbb Q_1(B)=1-\Lambda(x)$ and $z\geq y$, we have
\begin{align*}
\mathbb Q_2((X-x)\id_{B}+y\id_{B^c}\leq z)&=\frac{\mathbb P(\{X\leq z+x\}\cap B\cap B_2)+\mathbb P(B^c\cap B_2)}{\mathbb P(B_2)}\\
&=1-\frac{\mathbb P(\{X>z+x\}\cap B\cap B_2)}{\mathbb P(B_2)}.
\end{align*}
Observe that $$\mathbb P(\{X>z+x\}\cap A_x\cap B_2)=\mathbb P(\{X>z+x\}\cap B_1\cap B_2\cap\{U_X^{\mathbb P}\leq \alpha_x\})\leq \mathbb P(\{X>z+x\}\cap B_1\cap B_2\cap B_3)$$
for any $B_3\in\mathcal F$ such that $\mathbb P(B_1\cap B_2\cap B_3)\geq (1-\Lambda(x))\mathbb P(B_1)-\mathbb{P}(B_1\cap B_2^c)$. Note that  $\mathbb Q_1(B)=1-\Lambda(x)$ implies $\mathbb P(B_1\cap B_2\cap B)\geq (1-\Lambda(x))\mathbb P(B_1)-\mathbb{P}(B_1\cap B_2^c)$.  Hence we have $\mathbb P(\{X>z+x\}\cap A_x\cap B_2)\leq \mathbb P(\{X>z+x\}\cap B_1\cap B_2\cap B)\leq\mathbb P(\{X>z+x\}\cap B\cap B_2) $. This implies
$$\mathbb Q_2((X-x)\id_{B}+y\id_{B^c}\leq z)\leq \mathbb Q_2((X-x)\id_{A_x}+y\id_{A_x^c}\leq z),~z\geq y.$$ 

For $\X=L^\infty$, let $y<\essinf^{\mathbb Q_2}(X-x)$.
Then for $z<y$, clearly, $$\mathbb Q_2((X-x)\id_{B}+y\id_{B^c}\leq z)=\mathbb Q_2((X-x)\id_{A_x}+y\id_{A_x^c}\leq z)=0.$$ By the law-invariance and monotonicity of $\rho$, for $B\in\mathcal F$ satisfying $\mathbb Q_1(B)=1-\Lambda(x)$, and $y<\essinf^{\mathbb Q_2}(X-x)$, we have
 $$\rho^{\mathbb Q_2}((X-x)\id_{A_x}+y\id_{A_x^c})\leq \rho^{\mathbb Q_2}((X-x)\id_{B}+y\id_{B^c}).$$

\tbl{For the case that $\rho^{\mathbb Q_2}$ is an $\epsilon$-tail risk measure, let $y<\VaR_\epsilon^{\mathbb Q_2}(X-x)$. Then for $z<y$, $$\mathbb Q_2((X-x)\id_{B}+y\id_{B^c}\leq z)<1-\epsilon,$$
which implies $\VaR_\epsilon^{\mathbb Q_2}((X-x)\id_{B}+y\id_{B^c})\geq y$. Combing the above conclusion, we have
$$\frac{(\mathbb Q_2((X-x)\id_{B}+y\id_{B^c}\leq z)-(1-\epsilon))_+}{\epsilon}\leq \frac{(\mathbb Q_2((X-x)\id_{A_x}+y\id_{A_x^c}\leq z)-(1-\epsilon))_+}{\epsilon},~z\in\R.$$
Using the monotonicity of $\rho$ and the fact that $\rho$ is an $\epsilon$-tail risk measure, we have for $B\in\mathcal F$ satisfying $\mathbb Q_1(B)=1-\Lambda(x)$, and $y<\VaR_\epsilon^{\mathbb Q_2}(X-x)$,
 $$\rho^{\mathbb Q_2}((X-x)\id_{A_x}+y\id_{A_x^c})\leq \rho^{\mathbb Q_2}((X-x)\id_{B}+y\id_{B^c}).$$}
 Consequently,
 $$\inf_{y\in\R}\inf_{\mathbb Q_1(B)=1-\Lambda(x)}\left\{x+\rho^{\mathbb Q_2}((X-x)\id_{B}+y\id_{B^c})\right\}=\inf_{y\in\R}\left\{x+\rho^{\mathbb Q_2}((X-x)\id_{A_x}+y\id_{A_x^c})\right\}.$$
 Hence, \eqref{Minimizer11} holds.  The conclusion on optimal allocation follows from the same reasoning as the proof of Theorem \ref{Thone}. The details are omitted. We complete the proof. \qed

 {\bf Proof of Corollary \ref{Cor2}}.  
We first consider (i). Note that
$$x+\rho_g^{\mathbb Q_2}((X-x)\id_{A_x}+y\id_{A_x^c})=\rho_g^{\mathbb Q_2}(X\id_{A_x}+(x+y)\id_{A_x^c}).$$
If $\mathbb Q_1(B_2)\leq\lambda^+$, then there exists $x\in\R$ such that $\mathbb P(B_1\cap B_2^c)\geq (1-\Lambda(x))\mathbb P(B_1)$. Hence, $A_x=\emptyset$. By Proposition \ref{Pro:conditional}, we have
$\Lambda\VaR^{\mathbb Q_1}\dsquare \rho_g^{\mathbb Q_2}(X)=\inf_{x,y\in\R}\rho_g^{\mathbb Q_2}(x+y)=-\infty$.

Next, we consider the case $\mathbb Q_1(B_2)>\lambda^+$. \tbl{Suppose $y<\VaR_\epsilon^{\mathbb Q_2}(X)-x$}. For $z\geq x+y$, we have
\begin{equation}\label{eq:22}
\begin{aligned}
\mathbb{Q}_2(X\id_{A_x}+(x+y)\id_{A_x^c}\leq z)&=\frac{\mathbb P(A_x^c\cap B_2)+\mathbb P(\{X\leq z\}\cap A_x\cap B_2)}{\mathbb P(B_2)}\\
&=1-\beta_x+\frac{\mathbb P(\{X\leq z\}\cap B_1\cap B_2\cap\{U_X^{\mathbb P}\leq \alpha_x\})}{\mathbb P(B_2)},
\end{aligned}
\end{equation}
where $\beta_x=\frac{(1-\Lambda(x))\mathbb P(B_1)-\mathbb P(B_1\cap B_2^c)}{\mathbb P(B_2)}$. Moreover, \tbl{$\mathbb{Q}_2(X\id_{A_x}+(x+y)\id_{A_x^c}\leq z)<\epsilon$} for $z<x+y$. If $g(\beta)<1$, then there exists $x\in\R$ such that $g(\beta_x)<1$. By \eqref{eq:22}, we have  $\mathbb{Q}_2(X\id_{A_x}+(x+y)\id_{A_x^c}\leq z)\geq 1-\beta_x$ for all $z\geq x+y$, which implies $\VaR_{t}^{\mathbb Q_2}(X\id_{A_x}+(x+y)\id_{A_x^c})\leq x+y\to-\infty$ as $y\to -\infty$ for all $t\in [\beta_x, 1)$. Hence, for $x\in\R$ with $\beta_x=\beta$, we have
$$\inf_{y\in\R}\rho_g^{\mathbb Q_2}(X\id_{A_x}+(x+y)\id_{A_x^c})\leq \lim_{y\to-\infty}\int_{0}^{1}\VaR_t^{\mathbb Q_2}(X\id_{A_x}+(x+y)\id_{A_x^c})\d g(t)=-\infty.$$

Now, we assume $g(\beta)=1$.   Direct computation shows for $0<t<\beta_x\wedge \epsilon$, \begin{align}\label{eq:VaR}&\VaR_{t}^{\mathbb Q_2}(X\id_{A_x}+(x+y)\id_{A_x^c})\nonumber\\
&=\inf\left\{z\geq x+y: 1-\beta_x+\frac{\mathbb P(\{X\leq z\}\cap B_1\cap B_2\cap\{U_X^{\mathbb P}\leq \alpha_x\})}{\mathbb P(B_2)}
\geq 1-t\right\}\nonumber\\
&=\VaR_{1-\frac{\mathbb P(B_2)(\beta_x-t)}{\mathbb P(B_1\cap B_2)}}^{\mathbb Q}(X).
\end{align}
Hence, we have
\begin{align*}\Lambda\VaR^{\mathbb Q_1}\dsquare \rho_g^{\mathbb Q_2}(X)&=\inf_{x\in\R}\inf_{y\in\R}\int_{[0,\beta_x\wedge \epsilon)}\VaR_t^{\mathbb Q_2}(X\id_{A_x}+(x+y)\id_{A_x^c})\d g(t)\\
&=\inf_{x\in\R}\int_{[0,\beta_x\wedge \epsilon)}\VaR_{1-\frac{\mathbb P(B_2)(\beta_x-t)}{\mathbb P(B_1\cap B_2)}}^{\mathbb Q}(X)\d g(t)=\int_{[0,\beta)}\VaR_{1-\frac{\mathbb P(B_2)(\beta-t)}{\mathbb P(B_1\cap B_2)}}^{\mathbb Q}(X)\d g(t).
\end{align*}


Let us now consider case (ii). Similarly as above, if $\mathbb Q_1(B_2)\leq\lambda^+$, then  by Proposition \ref{Pro:conditional}, we have
$\Lambda\VaR^{\mathbb Q_1}\dsquare \Lambda_1\VaR^{+,\mathbb Q_2}(X)=\inf_{x,y\in\R}(x+y)=-\infty$.

Next, we consider the case $\mathbb Q_1(B_2)>\lambda^+$. In light of Proposition \ref{Pro:conditional} and \eqref{eq:22}, we have
\begin{align*}&\Lambda\VaR^{\mathbb Q_1}\dsquare \Lambda_1\VaR^{+,\mathbb Q_2}(X)\\
&=\inf_{x\in\R}\lim_{y\to -\infty}\left\{x+\sup\{z\geq y: 1-\beta_x+\frac{\mathbb P(\{X\leq x+z\}\cap B_1\cap B_2\cap\{U_X^{\mathbb P}\leq \alpha_x\})}{\mathbb P(B_2)}< 1-\Lambda_1(z) \}\right\}\\
&=\inf_{x\in\R}\inf_{y\to-\infty}\sup\left\{z\geq y+x: 1-\beta_x+\frac{\mathbb P(\{X\leq z\}\cap B_1\cap B_2\cap\{U_X^{\mathbb P}< \alpha_x\})}{\mathbb P(B_2)}< 1-\Lambda_1(z-x) \right\}\\
&=\inf_{x\in\R}\sup\left\{z\in\R: 1-\beta_x+\frac{\mathbb P(\{X\leq z\}\cap B_1\cap B_2)}{\mathbb P(B_2)}< 1-\Lambda_1(z-x) \right\}\\
&=\inf_{x\in\R}\sup\left\{z\in\R:\frac{\mathbb P(\{X\leq z\}\cap B_1\cap B_2)}{\mathbb P(B_2)}< \beta_x-\Lambda_1(z-x) \right\}\\
&=\inf_{x\in\R}\sup\left\{z\in\R:\frac{\mathbb P(\{X\leq z\}\cap B_1\cap B_2)}{\mathbb P(B_1\cap B_2)}< 1-\frac{\mathbb P(B_1)\Lambda(x)+\mathbb P(B_2)\Lambda_1(z-x)}{\mathbb P(B_1\cap B_2)} \right\}\\
&=\inf_{x\in\R}\sup\left\{z\in\R: F_{X}^{\mathbb Q}(z)< 1-\frac{\mathbb P(B_1)\Lambda(x)+\mathbb P(B_2)\Lambda_1(z-x)}{\mathbb P(B_1\cap B_2)} \right\}=\inf_{x\in\R}\overline{\Lambda}^x\VaR^{+,\mathbb Q}(X),
\end{align*}
where  $\overline{\Lambda}^x(z)=\frac{\mathbb P(B_1)\Lambda(x)+\mathbb P(B_2)\Lambda_1(z-x)}{\mathbb P(B_1\cap B_2)}\bigwedge 1$. We complete the proof. \qed

{\bf Proof of Corollary \ref{Cor3}}.  We first consider (i). In light of Theorem \ref{Thone}, we have
\begin{align*}\Lambda\VaR^{\mathbb Q_1}\dsquare \ES_{\alpha}^{\mathbb Q_2}(X)&=\inf_{x\in\R}\inf_{y\in\R}\inf_{\mathbb Q_1(B)=1-\Lambda(x)}\left\{x+\mathrm{ES}_{\alpha}^{\mathbb Q_2}((X-x)\id_{B}+y\id_{B^c})\right\}\\
&=\inf_{x\in\R}\inf_{y\in\R}\inf_{\mathbb Q_1(B)=1-\Lambda(x)}\left\{\mathrm{ES}_{\alpha}^{\mathbb Q_2}(X\id_{B}+(x+y)\id_{B^c})\right\}.
\end{align*}
It shows in \cite{RU02} that  $\ES_{\alpha}^{\mathbb Q_2}(X)=\inf_{t\in\R}\left(t+\frac{1}{\alpha}\mathbb E^{\mathbb Q_2}\left((X-t)_+\right)\right)$. Hence, we have
\begin{align*}\Lambda\VaR^{\mathbb Q_1}\dsquare \ES_{\alpha}^{\mathbb Q_2}(X)&=\inf_{x\in\R}\inf_{y\in\R}\inf_{\mathbb Q_1(B)=1-\Lambda(x)}\left\{\mathrm{ES}_{\alpha}^{\mathbb Q_2}(X\id_{B}+(x+y)\id_{B^c})\right\}\\
&=\inf_{x\in\R}\inf_{y\in\R}\inf_{\mathbb Q_1(B)=1-\Lambda(x)}\inf_{t\in\R}\left\{t+\frac{1}{\alpha}\mathbb E^{\mathbb Q_2}((X-t)_+\id_{B}+(x+y-t)_+\id_{B^c})\right\}\\
&=\inf_{x\in\R}\inf_{\mathbb Q_1(B)=1-\Lambda(x)}\inf_{t\in\R}\left\{t+\frac{1}{\alpha}\mathbb E^{\mathbb Q_2}((X-t)_+\id_B)\right\}\\
&=\inf_{x\in\R}\inf_{t\in\R}\inf_{\mathbb Q_1(B)=1-\Lambda(x)}\left\{t+\frac{1}{\alpha}\mathbb E^{\mathbb Q_1}(\eta(X-t)_+\id_{B})+\frac{1}{\alpha}\mathbb E^{\mathbb Q_2}((X-t)_+\id_{B\cap N})\right\}\\
&=\inf_{x\in\R}\inf_{t\in\R}\inf_{\mathbb Q_1(B)=1-\Lambda(x)}\left\{t+\frac{1}{\alpha}\mathbb E^{\mathbb Q_1}(\eta(X-t)_+\id_{B})\right\}\\
&=\inf_{x\in\R}\inf_{t\in\R}\left\{t+\frac{1}{\alpha}\int_{\Lambda(x)}^{1}\VaR_s^{\mathbb Q_1}(\eta(X-t)_+)\d s\right\}\\
&=\inf_{t\in\R}\left\{t+\frac{1}{\alpha}\int_{\lambda^+}^{1}\VaR_s^{\mathbb Q_1}(\eta(X-t)_+)\d s\right\}.
\end{align*}

Next, we consider (ii). It follows from Theorem \ref{Thone} that
\begin{align*}\Lambda\VaR^{\mathbb Q_1}\dsquare \mathbb E_u^{\mathbb Q_2}(X)&=\inf_{x\in\R}\inf_{y\in\R}\inf_{\mathbb Q_1(B)=1-\Lambda(x)}\left\{x+\mathbb E^{\mathbb Q_2}\left(u(X-x)\id_{B}+u(y)\id_{B^c}\right)\right\}\\
&=\inf_{x\in\R}\inf_{\mathbb Q_1(B)=1-\Lambda(x)}\left\{x+\mathbb E^{\mathbb Q_2}\left(u(X-x)\id_{B}+u(-\infty)\id_{B^c}\right)\right\}.
\end{align*}
Note that $u(-\infty)\leq u(X-x)$. Hence, we have
\begin{align*}&\Lambda\VaR^{\mathbb Q_1}\dsquare \mathbb E_u^{\mathbb Q_2}(X)\\
&=\inf_{x\in\R}\inf_{\mathbb Q_1(B)=1-\Lambda(x)}\left\{x+\mathbb E^{\mathbb Q_1}(\eta u(X-x)\id_{B})+u(-\infty)\mathbb E^{\mathbb Q_1}(\eta\id_{B^c})\right.\\
&~~~~\left.+\mathbb E^{\mathbb Q_2}(u(X-x)\id_{B\cap N})+u(-\infty)\mathbb E^{\mathbb Q_2}(\id_{B^c\cap N})\right\}\\
&=\inf_{x\in\R}\left\{x+\mathbb E^{\mathbb Q_1}(\eta u(X-x)\id_{\{U_\eta^{\mathbb Q_1}<1-\Lambda(x)\}})+u(-\infty)\left(\mathbb Q_2(N)+\int_{0}^{\Lambda(x)}\VaR_t^{\mathbb Q_1}(\eta)\d t\right)\right\}.
\end{align*}
\qed

{\bf Proof of Theorem \ref{Thfour}}. For $x\in\R, y\geq x$ and $U\overset{\mathbb Q_1}{\sim} U[0,1]$, we have $\Lambda\VaR^{+,\mathbb Q_1}(\Lambda_{x,y}^{-1}(U))=x$. Hence,
$$\Lambda\VaR^{+,\mathbb Q_1}\dsquare \rho(X)\leq \Lambda\VaR^{+,\mathbb Q_1}(\Lambda_{x,y}^{-1}(U))+\rho^{\mathbb Q_2}(X-\Lambda_{x,y}^{-1}(U))=x+\rho^{\mathbb Q_2}(X-\Lambda_{x,y}^{-1}(U)).$$
Using the arbitrary of $x\in\R, y\geq x$ and $U\overset{\mathbb Q_1}{\sim} U[0,1]$, we have
$$\Lambda\VaR^{+,\mathbb Q_1}\dsquare \rho^{\mathbb Q_2}(X)\leq \inf_{x\in\R}\inf_{y\geq x}\inf_{U\overset{\mathbb Q_1}{\sim} U[0,1]}\left\{x+\rho\left(X-\Lambda_{x,y}^{-1}(U)\right)\right\}.$$
Next, we show the inverse inequality.  For $X_1\in\X$, we let $x_1=\Lambda\VaR^{+,\mathbb Q_1}(X_1)$. We choose $U_{X_1}^{\mathbb Q_1}$  such that $\{U_{X_1}^{\mathbb Q_1}\leq t\}\subset\{X_1\leq F_{X_1}^{\mathbb Q_1, -1}(t)\}$ for all $t\in (0,1)$. This implies $\{F_{X_1}^{\mathbb Q_1, -1}(U_{X_1}^{\mathbb Q_1})\leq x\}=\{U_{X_1}^{\mathbb Q_1}\leq F_{X_1}^{\mathbb Q_1}(x)\}\subset\{X_1\leq F_{X_1}^{\mathbb Q_1, -1}(F_{X_1}^{\mathbb Q_1}(x))\}\subset\{X_1\leq x\}$ for all $x\in\R$. Hence, $X_1\leq F_{X_1}^{\mathbb Q_1, -1}(U_{X_1}^{\mathbb Q_1})$ for all $w\in\Omega$.  By definition, $F_{X_1}^{\mathbb Q_1}(x)\geq 1-\Lambda(x)$ for all $x\geq x_1$. This implies $F_{X_1}^{\mathbb Q_1}(x)\geq \Lambda_{x_1,y}(x)$ for $x_1\leq x\leq y$.  Suppose $\esssup^{\mathbb Q_1} X_1<\infty$. If $y\geq \esssup^{\mathbb Q_1} X_1$, then $F_{X_1}^{\mathbb Q_1}(x)=1\geq \Lambda_{x_1,y}(x)$ for $x\geq y$. Hence, if $y\geq \esssup^{\mathbb Q_1} X_1$, we have $F_{X_1}^{\mathbb Q_1}(x)\geq \Lambda_{x_1,y}(x)$ for all $x\in\R$. Combing all the above results, we have $X_1\leq F_{X_1}^{\mathbb Q_1, -1}(U_{X_1}^{\mathbb Q_1})\leq \Lambda_{x_1,y}^{-1}(U_{X_1}^{\mathbb Q_1})$ if $y\geq \esssup^{\mathbb Q_1} X_1$. It follows that for $y\geq \esssup^{\mathbb Q_1} X_1$,
\begin{align*}\Lambda\VaR^{+,\mathbb Q_1}(X_1)+\rho^{\mathbb Q_2}(X-X_1)&\geq x_1+\rho^{\mathbb Q_2}(X-\Lambda_{x_1,y}^{-1}(U_{X_1}^{\mathbb Q_1}))\\
&\geq \inf_{x\in\R}\inf_{y\geq x\vee \esssup^{\mathbb Q_1} X_1}\inf_{U\overset{\mathbb Q_1}{\sim} U[0,1]}\left\{x+\rho^{\mathbb Q_2}\left(X-\Lambda_{x,y}^{-1}(U)\right)\right\}\\
&=\inf_{x\in\R}\inf_{y\geq x}\inf_{U\overset{\mathbb Q_1}{\sim} U[0,1]}\left\{x+\rho^{\mathbb Q_2}\left(X-\Lambda_{x,y}^{-1}(U)\right)\right\}.
\end{align*}
\tbl{If $\esssup^{\mathbb Q_1} X_1=\infty$, then let $X_m=X_1\wedge m$ for $m\geq 2$. Using the above conclusion, we have 
\begin{align*}\Lambda\VaR^{+,\mathbb Q_1}(X_m)+\rho^{\mathbb Q_2}(X-X_m)\geq \inf_{x\in\R}\inf_{y\geq x}\inf_{U\overset{\mathbb Q_1}{\sim} U[0,1]}\left\{x+\rho^{\mathbb Q_2}\left(X-\Lambda_{x,y}^{-1}(U)\right)\right\}.
\end{align*}
Direct computation gives $\lim_{m\to\infty} \Lambda\VaR^{+,\mathbb Q_1}(X_m)=\Lambda\VaR^{+,\mathbb Q_1}(X_1)$. Note that $X-X_m\downarrow X-X_1$ as  $m\to\infty$. Using the fact that $\rho$ is continuous from above, we have $\lim_{m\to\infty}\rho^{\mathbb Q_2}(X-X_m)=\rho^{\mathbb Q_2}(X-X_1)$. Hence, letting $m\to\infty$, we obtain
\begin{align*}\Lambda\VaR^{+,\mathbb Q_1}(X_1)+\rho^{\mathbb Q_2}(X-X_1)\geq \inf_{x\in\R}\inf_{y\geq x}\inf_{U\overset{\mathbb Q_1}{\sim} U[0,1]}\left\{x+\rho^{\mathbb Q_2}\left(X-\Lambda_{x,y}^{-1}(U)\right)\right\}.
\end{align*}}
By the arbitrary of $X_1$, we have $$\Lambda\VaR^{+,\mathbb Q_1}\dsquare \rho^{\mathbb Q_2}(X)\geq \inf_{x\in\R}\inf_{y\geq x}\inf_{U\overset{\mathbb Q_1}{\sim} U[0,1]}\left\{x+\rho^{\mathbb Q_2}\left(X-\Lambda_{x,y}^{-1}(U)\right)\right\}.$$  Hence, we obtain \eqref{Minimizer2}.

Next, we show for $\X=L^\infty$, the optimal allocation of the inf-convolution exists if and only if the minimizer of \eqref{Minimizer2} exists. Suppose $(X_1, X-X_1)$ is the optimal allocation of the inf-convolution. Then we have $\Lambda\VaR^{+,\mathbb Q_1}(X_1)+\rho^{\mathbb Q_2}(X-X_1)=\Lambda\VaR^{+,\mathbb Q_1}\square\rho^{\mathbb Q_2}(X)$. Let $x_1=\Lambda\VaR^{+,\mathbb Q_1}(X_1)$ and $y_1>x_1\vee\esssup^{\mathbb Q_1} X_1$. We choose $U_{X_1}^{\mathbb Q_1}$ such that $X_1\leq F_{X_1}^{\mathbb Q_1, -1}(U_{X_1}^{\mathbb Q_1})$ for all $w\in\Omega$. Using the above argument, we have $F_{X_1}^{\mathbb Q_1, -1}(U_{X_1}^{\mathbb Q_1})\leq \Lambda_{x_1,y_1}^{-1}(U_{X_1}^{\mathbb Q_1})$. By the monotonicity of $\rho$, we have $\Lambda\VaR^{+,\mathbb Q_1}(X_1)+\rho^{\mathbb Q_2}(X-X_1)\geq x_1+\rho^{\mathbb Q_2}(X-\Lambda_{x_1,y_1}^{-1}(U_{X_1}^{\mathbb Q_1}))$. This implies $x_1+\rho^{\mathbb Q_2}(X-\Lambda_{x_1,y_1}^{-1}(U_{X_1}^{\mathbb Q_1}))=\inf_{x\in\R}\inf_{y\geq x}\inf_{U\overset{\mathbb Q_1}{\sim} U[0,1]}\left\{x+\rho^{\mathbb Q_2}\left(X-\Lambda_{x,y}^{-1}(U)\right)\right\}$. Hence, $(x_1,y_1,U_{X_1}^{\mathbb Q_1})$ is the minimizer of \eqref{Minimizer2}.
Moreover, if $(x_1,y_1,U)$ is the minimizer of \eqref{Minimizer2}, one can easily check that $(\Lambda_{x_1,y_1}^{-1}(U), X-\Lambda_{x_1,y_1}^{-1}(U))$ is the optimal allocation of the inf-convolution. We complete the proof. \qed

{\bf Proof of Corollary \ref{Cor:-1}}. \tbl{In light of Theorem \ref{Thfour}, we have
\begin{align*}\Lambda\VaR^{+,\mathbb Q}\dsquare \overline{\Lambda_1\VaR}^{\mathbb Q} (X)=\inf_{x\in\R}\inf_{y\geq x}\inf_{U\overset{\mathbb Q}{\sim} U[0,1]}\left\{x+\overline{\Lambda_1\VaR}^{\mathbb Q}\left(X-\Lambda_{x,y}^{-1}(U)\right)\right\}.
\end{align*}
Let $\hat{F}_{x,y}(z)=\{F^{\mathbb Q}_{X-\Lambda_{x,y}^{-1}(U)}(z): U\overset{\mathbb Q}{\sim} U[0,1]\}$ for $z\in \R$ and $\overline{\VaR}_{\beta}^{\mathbb Q}$ represent $\overline{\Lambda_1\VaR}^{\mathbb Q}$ for $\Lambda_1=\beta\in (0,1)$. This means 
$\overline{\VaR}_{\beta}^{\mathbb Q}$ is the right $\VaR$.
By Theorem 1 of \cite{HL24}, we have
$$\inf_{U\overset{\mathbb Q}{\sim} U[0,1]}\overline{\Lambda_1\VaR}^{\mathbb Q}\left(X-\Lambda_{x,y}^{-1}(U)\right)=q_{\Lambda_1}^+(\hat{F}_{x,y}).$$
It follows from Corollary 1 of \cite{HL24} that for $\alpha\in (0,1)$, $$q_{\alpha}^+(\hat{F}_{x,y})=\inf_{U\overset{\mathbb Q}{\sim} U[0,1]} \overline{\VaR}_{\alpha}^{\mathbb Q}(X-\Lambda_{x,y}^{-1}(U)),$$
where $q_{\alpha}^+$ is $q_{\Lambda_1}^+$ with $\Lambda_1=\alpha$.
Direct computation shows 
$$\inf_{U\overset{\mathbb Q}{\sim} U[0,1]} \overline{\VaR}_{\alpha}^{\mathbb Q}(X-\Lambda_{x,y}^{-1}(U))=-\sup_{U\overset{\mathbb Q}{\sim} U[0,1]} \VaR_{1-\alpha}^{\mathbb Q}(-X+\Lambda_{x,y}^{-1}(U)).$$
In light of Lemma 4.3 of \cite{BJW14} and the result in \cite{M81}, we have
\begin{align*}\sup_{Y\overset{\mathbb Q}{\sim} F_X^{\mathbb Q}, U\overset{\mathbb Q}{\sim} U[0,1]} \overline{\VaR}_{1-\alpha}^{\mathbb Q}(-Y+\Lambda_{x,y}^{-1}(U))&=\overline{\VaR}_{1-\alpha}^{\mathbb Q}(-X+\Lambda_{x,y}^{-1}(V))\\
&\leq \sup_{U\overset{\mathbb Q}{\sim} U[0,1]} \overline{\VaR}_{1-\alpha}^{\mathbb Q}(-X+\Lambda_{x,y}^{-1}(U)),
\end{align*}
where $V=U_{-X}^{\mathbb Q}\id_{\{U_{-X}^{\mathbb Q}\leq \alpha\}}+(1+\alpha-U_{-X}^{\mathbb Q})\id_{\{U_{-X}^{\mathbb Q}>\alpha\}}$,
which implies 
\begin{align*}\sup_{U\overset{\mathbb Q}{\sim} U[0,1]} \overline{\VaR}_{1-\alpha}^{\mathbb Q}(-X+\Lambda_{x,y}^{-1}(U))&=\sup_{Y\overset{\mathbb Q}{\sim} F_X^{\mathbb Q}, U\overset{\mathbb Q}{\sim} U[0,1]} \overline{\VaR}_{1-\alpha}^{\mathbb Q}(-Y+\Lambda_{x,y}^{-1}(U))\\
&=\overline{\VaR}_{1-\alpha}^{\mathbb Q}(-X+\Lambda_{x,y}^{-1}(V)).
\end{align*}
Using the above conclusion and the continuity of $\Lambda_{x,y}^{-1}$ and $F_X^{\mathbb Q, -1}$, applying Lemma 4.4 of \cite{BJW14}, we have
$$\sup_{U\overset{\mathbb Q}{\sim} U[0,1]} \VaR_{1-\alpha}^{\mathbb Q}(-X+\Lambda_{x,y}^{-1}(U))=\sup_{U\overset{\mathbb Q}{\sim} U[0,1]} \overline{\VaR}_{1-\alpha}^{\mathbb Q}(-X+\Lambda_{x,y}^{-1}(U)).$$
Hence,
$$q_{\alpha}^+(\hat{F}_{x,y})=-\overline{\VaR}_{1-\alpha}^{\mathbb Q}(-X+\Lambda_{x,y}^{-1}(V))=l_{x,y}(1-\alpha),$$
where $l_{x,y}(t)=\sup_{s\in (0,t]}\{F_X^{\mathbb Q,-1}(s)-\Lambda_{x,y}^{-1}(1-t+s)\},~t\in (0,1)$.
Direct computation shows 
$\hat{F}_{x,y}(z)=l_{x,y}^{-1}(z)$  on $\R$ except a countable number of points. }

\tbl{
Next, we show $q_{\Lambda_1}^+(l_{x,y}^{-1})=q_{\Lambda_1}^+(\hat{F}_{x,y})$. Let $z_0=q_{\Lambda_1}^+(\hat{F}_{x,y})$. Clearly, $z_0\in \R$. By definition, there exist $z_n\downarrow z_0$ such that $\hat{F}_{x,y}(z_n)>1-\Lambda_1(z_n)$. It follows from the right continuity of $\Lambda_1$ and the monotonicity of $\hat{F}_{x,y}$ that there exist $0<\epsilon_n<1/n$ such that $\hat{F}_{x,y}(z)>1-\Lambda_1(z)$ for $z\in (z_n,z_n+\epsilon_n)$. Then there exist $w_n\in (z_n,z_n+\epsilon_n)$ such that $l_{x,y}^{-1}(w_n)=\hat{F}_{x,y}(w_n)>1-\Lambda_1(w_n)$. Hence, $q_{\Lambda_1}^+(l_{x,y}^{-1})\leq w_n\to z_0=q_{\Lambda_1}^+(\hat{F}_{x,y})$. We can similarly show the inverse inequality. Hence,   $q_{\Lambda_1}^+(l_{x,y}^{-1})=q_{\Lambda_1}^+(\hat{F}_{x,y})$. We complete the proof. \qed}

{\bf Proof of Proposition \ref{prop:step}}. Direct calculation shows that if $x<x_1$, then
$\Lambda_{x,y}(z)=\lambda_1\id_{\{x\leq z<x_1\}}+\lambda_2\id_{\{x_1\leq z<y\}}+\id_{\{z\geq y\}}$; if $x\geq x_1$, then
$\Lambda_{x,y}(z)=\lambda_2\id_{\{x\leq z<y\}}+\id_{\{z\geq y\}}$.
Hence, we have $\Lambda_{x,y}^{-1}(t)=x\id_{\{0<t\leq \lambda_1\}}+x\vee x_1\id_{\{\lambda_1<t\leq \lambda_2\}}+y\id_{\{\lambda_2<t<1\}}$.
For any $U\overset{\mathbb Q_1}{\sim} U[0,1]$, it follows that $\Lambda_{x,y}^{-1}(U)=x\id_{\{0<U\leq \lambda_1\}}+x\vee x_1\id_{\{\lambda_1<U\leq \lambda_2\}}+y\id_{\{\lambda_2<U<1\}}$. Note that $\mathbb Q_1(0<U\leq \lambda_1)=\lambda_1$, $\mathbb Q_1(\lambda_1<U\leq \lambda_2)=\lambda_2-\lambda_1$ and $\mathbb Q_1(\lambda_2<U<1)=1-\lambda_2$, and those three sets are disjoint. Consequently, in light of Theorem \ref{Thfour}, we have
\begin{align*}
\Lambda\VaR^{+,\mathbb Q_1}\dsquare \rho^{\mathbb Q_2}(X)&=\inf_{x\in\R}\inf_{y\geq x}\inf_{U\overset{\mathbb Q_1}{\sim} U[0,1]}\left\{x+\rho^{\mathbb Q_2}\left(X-\Lambda_{x,y}^{-1}(U)\right)\right\}\\
&=\inf_{x\in\R}\inf_{y\geq x\vee x_1}\inf_{(B_1,B_2)\in\mathcal B_{\lambda_1,\lambda_2}}\left\{x+\rho^{\mathbb Q_2}\left(X-x\id_{B_1}-x\vee x_1\id_{B_2}-y\id_{(B_1\cup B_2)^c}\right)\right\}.
\end{align*} \qed

\tbl{
{\bf Proof of Corollary \ref{Cor:-2}}. In light of Proposition \ref{prop:step}, we have
\begin{align*}
\Lambda\VaR^{+,\mathbb Q_1}\dsquare \ES_{\alpha}^{\mathbb Q_2}(X)&=\inf_{x\in\R}\inf_{y\geq x\vee x_1}\inf_{(B_1,B_2)\in\mathcal B_{\lambda_1,\lambda_2}}\left\{x+\ES_{\alpha}^{\mathbb Q_2}\left(X-x\id_{B_1}-x\vee x_1\id_{B_2}-y\id_{(B_1\cup B_2)^c}\right)\right\}\\
&=\inf_{x\in\R}\inf_{y\geq x\vee x_1}\inf_{(B_1,B_2)\in\mathcal B_{\lambda_1,\lambda_2}}\ES_{\alpha}^{\mathbb Q_2}\left(X+0\wedge (x-x_1)\id_{B_2}+(x-y)\id_{(B_1\cup B_2)^c}\right)\\
&=\inf_{(B_1,B_2)\in\mathcal B_{\lambda_1,\lambda_2}}\lim_{y\to\infty}\ES_{\alpha}^{\mathbb Q_2}\left(X-y\id_{B_1^c}\right).
\end{align*}
From the above expression, we should choose $B_1$ such that $\mathbb Q_2(B_1)$ is as small as possible. Hence, if $\lambda_1\mathbb P(D_1)-\mathbb P(D_1\cap D_2^c)\geq \alpha\mathbb P(D_2)$, the optimal way to choose $B_1$ is to set $B_1=(D_1\cap D_2^c)\cup(\{U_X^{\mathbb P}<t^+\}\cap D_1\cap D_2)$  with $\mathbb P(\{U_X^{\mathbb P}<t^+\}\cap D_1\cap D_2)=\lambda_1\mathbb P(D_1)-\mathbb P(D_1\cap D_2^c)$.
Then we have $$\Lambda\VaR^{+,\mathbb Q_1}\dsquare \ES_{\alpha}^{\mathbb Q_2}(X)=\lim_{y\to\infty}\ES_{\alpha}^{\mathbb Q_2}\left(X-y\id_{B_1^c}\right)=\frac{1}{\alpha}\mathbb E^{\mathbb Q_2}\left(X\id_{\{t^-<U_X^{\mathbb P}<t^+\}\cap D_1\cap D_2}\right)$$
with $\mathbb P(\{t^-<U_X^{\mathbb P}<t^+\}\cap D_1\cap D_2)=\alpha \mathbb P(D_2)$. 
Note that $\frac{d\mathbb Q_2}{d\mathbb Q}=\frac{\mathbb P(D_1\cap D_2)}{\mathbb P(D_2)}$ on $D_1\cap D_2$. Hence, we have
\begin{align*}\mathbb E^{\mathbb Q_2}\left(X\id_{\{t^-<U_X^{\mathbb P}<t^+\}\cap D_1\cap D_2}\right)&=\frac{\mathbb P(D_1\cap D_2)}{\mathbb P(D_2)}\mathbb E^{\mathbb Q}\left(X\id_{\{t^-<U_X^{\mathbb P}<t^+\}\cap D_1\cap D_2}\right)\\
&=\frac{\mathbb P(D_1\cap D_2)}{\mathbb P(D_2)}\int_{\alpha_0}^{\alpha_1}\VaR_{t}^{\mathbb Q}(X)\d t.
\end{align*}
with $\alpha_0=1-\frac{\lambda_1\mathbb P(D_1)-\mathbb P(D_1\cap D_2^c)}{\mathbb P(D_1\cap D_2)}$ and $\alpha_1=1-\frac{\lambda_1\mathbb P(D_1)-\mathbb P(D_1\cap D_2^c)-\alpha\mathbb P(D_2)}{\mathbb P(D_1\cap D_2)}$, which implies the desired result.}

\tbl{
If $\lambda_1\mathbb P(D_1)-\mathbb P(D_1\cap D_2^c)\leq 0$, we choose $B_1\subset D_1\cap D_2^c$. Then $\mathbb Q_2(B_1^c)=1$, which implies $\Lambda\VaR^{\mathbb Q_1}\dsquare \ES_{\alpha}^{\mathbb Q_2}(X)=-\infty$.}

\tbl{
If $0<\lambda_1\mathbb P(D_1)-\mathbb P(D_1\cap D_2^c)<\alpha\mathbb P(D_2)$, then we choose $B_1=(D_1\cap D_2^c)\cup(\{U_X^{\mathbb P}<t^+\}\cap D_1\cap D_2)$, where $t^+$ is defined above. Direct computation shows $\mathbb Q_2(B_1)<\alpha$, which implies 
$$\Lambda\VaR^{+,\mathbb Q_1}\dsquare \ES_{\alpha}^{\mathbb Q_2}(X)=\lim_{y\to\infty}\ES_{\alpha}^{\mathbb Q_2}\left(X-y\id_{B_1^c}\right)=-\infty.$$ 
This completes the proof. \qed}
\end{document}